\documentclass[usenatbib]{mn2e}

\usepackage{graphicx}
\usepackage{url}
\usepackage{amsmath}
\usepackage{threeparttable}
\usepackage{url}
\usepackage{hyperref}
\usepackage{multirow}
\usepackage{caption3}
\usepackage[all]{hypcap}        
\usepackage[capitalise]{cleveref}
\crefformat{equation}{Eq.~#2#1#3}

\usepackage{xcolor} 
\definecolor{darkblue}{rgb}{0,0,0.5}
\hypersetup{
  colorlinks   = true, 
  urlcolor     = blue, 
  linkcolor    = blue, 
  citecolor    = darkblue   
}

\graphicspath{{figures/}} 

\newcommand{\msun}{\mbox{\,M$_\odot$}}        
\newcommand{\kms}{km~s$^{-1}$}
\newcommand{\Ha}{H$\alpha$}
\newcommand{\vrec}{$v_\mathrm{rec}$}
\newcommand{\ca}{Ca-rich SN}
\newcommand{\cas}{Ca-rich SNe}

\newcommand{\simlt}{{\small\raisebox{-0.6ex}{$\,\stackrel{\raisebox{-.2ex}{$\textstyle <$}}{\sim}\,$}}}
\newcommand{\simgt}{{\small\raisebox{-0.6ex}{$\,\stackrel{\raisebox{-.2ex}{$\textstyle >$}}{\sim}\,$}}}
\newcommand{\hst}{{\em HST}}
\newcommand{\sex}{{\sc SExtractor}}

\begin{document}

\title[Environments of Ca-rich SNe]{{\em Hubble Space Telescope} observations of the host galaxies and environments of calcium-rich supernovae}

\author[Lyman et al.]
{\parbox{\textwidth}{J. D. Lyman$^1$\thanks{E-mail: J.D.Lyman@warwick.ac.uk}, 
A. J. Levan$^1$,
P. A. James$^2$,
C. R. Angus$^1$,
R. P. Church$^3$,
M. B. Davies$^3$,
N. R. Tanvir$^4$
}
\vspace{0.4cm}\\
$^1$Department of Physics, University of Warwick, Coventry CV4 7AL, UK\\
$^2$Astrophysics Research Institute, Liverpool John Moores University, Liverpool, L3 5RF, UK\\
$^3$Lund Observatory, Department of Astronomy and Theoretical Physics, Box 43, SE-221 00 Lund, Sweden\\
$^4$Department of Physics and Astronomy, University of Leicester, Leicester, LE1 7RH, UK\\
}

\date{}

\pagerange{\pageref{firstpage}--\pageref{lastpage}} \pubyear{2015}

\maketitle

\label{firstpage}

\begin{abstract}
Calcium-rich supernovae represent a significant challenge for our understanding of the fates of stellar systems. They are less luminous than other supernova (SN) types and they evolve more rapidly to reveal nebular spectra dominated by strong calcium lines with weak or absent signatures of other intermediate- and iron-group elements, which are seen in other SNe. Strikingly, their explosion sites also mark them out as distinct from other SN types. Their galactocentric offset distribution is strongly skewed to very large offsets ($\sim1/3$ are offset $>$20~kpc), meaning they do not trace the stellar light of their hosts. Many of the suggestions to explain this extreme offset distribution have invoked the necessity for unusual formation sites such as globular clusters or dwarf satellite galaxies, which are therefore difficult to detect. Building on previous work attempting to detect host systems of nearby \cas{}, we here present {\em Hubble Space Telescope} imaging of 5 members of the class -- 3 exhibiting large offsets and 2 coincident with the disk of their hosts. We find no underlying sources at the explosion sites of any of our sample. Combining with previous work, the lack of a host system now appears to be a ubiquitous feature amongst \cas{}. In this case the offset distribution is most readily explained as a signature of high-velocity progenitor systems that have travelled significant distances before exploding.

\end{abstract}


\section{Introduction}
\label{sect:intro}

Calcium-rich supernovae (Ca-rich SNe) are a recently defined breed of `fast-and-faint' transient whose observational properties prove challenging to explain. Following an initial investigation of the prototypical SN~2005E \citep{perets10}, \citet{kasliwal12} detailed the characteristics of this class that makes them distinct from more well known SN types. Their light curves exhibit peak luminosities intermediate between novae and supernovae, and evolve more quickly than typical SNe. Peak-light spectra have resemblances to Type Ib SNe (SNe~Ib, i.e. a lack of detected hydrogen and conspicuous helium, although see \citealt{sullivan11, valenti14}), and also have photospheric velocities typical of other SN types, i.e.\ $\sim$10000~\kms{}. However, their ejecta quickly evolve to the optically-thin nebular phase ($\sim 90$d), where they exhibit further differences from other SNe. Optical nebular spectra are dominated by [Ca~{\sc ii}], with very little, if any, evidence of other intermediate mass or Fe-group elements, which show strong features in core-collapse and thermonuclear SN~Ia spectra. Most peculiarly, \cas{} do not trace the stellar light of their hosts, unlike other luminous transients and SNe \citep{anderson15}. A large fraction are found at significant distances from the nearest galaxy, well outside the bulk of the stellar light, raising further questions about the nature of their progenitors. 

Their similarity around peak light to SNe~Ib (expected to arise from progenitors more massive than 8\msun{}) led to some claims that \cas{} may also be due to the collapse of massive stars. This scenario, however, would be very surprising given the prevalence ($\sim$50~per~cent) of early-type host galaxies \citep{perets10, lyman13}, the lack of evidence for star formation or other massive stars at their locations \citep{perets10, perets11, lyman13, lyman14d}, and their extreme galactocentric offset distribution \citep{kasliwal12, yuan13, foley15}.

Although \cas{} appear much fainter and evolve more quickly than normal SNe~Ia, and thus the detonation of a Chandrasekhar mass white dwarf (WD) is an unlikely explanation for their formation, the presence of a WD in the progenitor systems of \cas{} is an attractive possibility. Not least, an explosion involving an older WD-related system is consistent with their lack of association to recent star-formation and the large fraction of early type hosts. The nature of the explosion mechanism of such a putative WD has been variously proposed as helium shell detonations on accreting WDs \citep{shen09,perets10,waldman11,meng15}, mergers of WD binaries \citep[e.g.][]{brown11,foley15}, mergers of WD-NS binaries \citep{metzger12}, and tidal disruptions and detonations of WDs by black holes \citep{sell15,macleod15}.

Their apparent preference for remote locations and large galactocentric offsets differentiates them from other SNe and luminous transients, and is another trait that must be explained by progenitor models. Five of the thirteen members of the class appear  further than 20~kpc from the centres of their respective hosts in plane of sky \citep{foley15}.\footnote{Alternatively expressed, these 5 examples are all offset  greater than six isophotal radii ($K_s = 20$~mag~arcsec$^{-2}$) from their hosts.} Preferential formation of the progenitors in faint dwarf galaxies or globular clusters (GCs) could provide an explanation for this preference for large offsets \citep{kasliwal12,yuan13}. The limits of \citet{kasliwal12} at the locations of three more distant examples ruled out only bright ($M_\text{R} \simlt -10$ to $-12$~mag) dwarf satellite galaxies. \citet{perets11} presented {\em Hubble Space Telescope} (\hst{}) imaging for SN~2005cz and found no evidence of a post-explosion source at its location, ruling out massive stars and brighter globular clusters.
To further investigate the presence of underlying host systems, \citet{lyman14d} presented deep Very Large Telescope (VLT) imaging at the location of two nearby members, SNe 2005E and 2012hn. These data probed down to limits of even faint globular clusters and dwarf galaxies ($M_\text{R} \sim -5$~mag) at the distance of the host galaxies. The lack of any detections ruled out the presence of underlying GCs or dwarf galaxies at the explosion sites of these two nearby examples.\footnote{An analysis of archival \hst{} data of SN~2003H in \citet{lyman14d} also found no distinct underlying source for this event, although limits were shallower owing to the explosion site being located within the disk of the host galaxy system.} 

The lack of underlying host systems is problematic for most WD progenitors -- such progenitors would be expected to trace the stellar light of their host systems, as is seen for SNe~Ia \citep[e.g.][]{forster08}. By contrast, all four of the \cas{} discovered by the Palomar Transient Factory \citep{law09} are at a galactocentric offset greater than $\sim 99$~per~cent of the SNe~Ia detected by the survey \citep{kasliwal12, foley15}. Combining their limits with previous literature limits, \citet{lyman14d} suggested these results indicate that the progenitors of \cas{} are not formed at these large offsets, but instead travel there as high-velocity systems, with kicked WD-NS binaries suggested as a potential progenitor system.
This suggestion was supported by an investigation of the kinematics of the transients relative to their hosts by \citet{foley15}. The velocity offsets of the transients around their hosts appeared inconsistent with those expected for bound-orbit systems, and instead pointed to the progenitors being expelled from their hosts. \citet{foley15} proposed a progenitor model of merging WD-WD binaries which have undergone a supermassive black hole interaction. As well as imparting a high velocity to the system, this interaction acts to harden the binaries, causing the merging timescale to decrease. This decrease in the merging timescale would need to be a significant factor in their production, in order that the rate of unperturbed disk WD-WD systems does not dominate over the high-velocity subset. An additional prediction of this model is a high rate of events close to the nuclear regions of galaxies. Although \cas{} are relatively low-luminosity, they have been detected in the bright disks of nearby galaxies. However, detecting transients coincident with the very bright central regions of galaxies in transient surveys is more difficult owing to the much brighter background and image subtraction residuals.

As further noted in \citet{foley15}, the rate of ongoing or recent merging or disturbance of the host galaxies is high, with 7/13 hosts show evidence of being the result of recent merging activity. Additionally, a number of the host galaxies show indirect evidence for interaction and disturbance based on the dense environments (groups/clusters) in which they reside. The number of hosts with either direct or environmental evidence for recent interaction is then 11/13. 

In this paper we present new \hst{} imaging of the hosts and explosion locations of five \cas{}. These are used to further investigate the nature of the hosts themselves, and attempt to detect the presence of any underlying host systems at the explosion sites of the \cas{}. The use of \hst{} imaging allows us to probe to the faint end of the luminosity function of dwarf galaxies and GCs even in these more distant examples \citep[cf.\ ][]{lyman14d}. This will allow us to test these as the locations of the \ca{} explosions, thereby increasing the number of examples with deep imaging in order to rule on the \ca{} class as a whole. In \cref{sect:obs} we present our sample and data reduction, and detail our analysis methods in \cref{sect:analysis}. Results are presented and discussed in \cref{sect:results,sect:discuss}, respectively. A summary is provided in \cref{sect:summary}.

\section{Sample and observations}
\label{sect:obs}

Our sample consists of five examples of the \ca{} class, which were targeted with \hst{} for two orbits each (GO13698, PI: Lyman). Throughout the paper, the description `remote' is used to describe SNe~2003dr, 2005E and 2007ke within our sample (i.e.\ those outside the bulk of the stellar light of their hosts). We will refer to SNe~2001co and 2003dg as `disk' transients since, in projection, they are located within the stellar disk of their hosts. These are simply empirical distinctions to distinguish our different analysis and observing strategies for each.
For the three remote examples, deep imaging with the Advanced Camera for Surveys/Wide Field Channel (ACS/WFC) was obtained in two wide filters, F435W and F606W, for one orbit each. For the two disk examples, Wide Field Camera 3 (WFC3)/UVIS was used to take wide filter imaging in F438W and F606W for half an orbit each and a narrowband F665N (filter transmission $\sim 6600-6700$\AA{}) image for the second orbit. In order to produce \Ha{} maps of the hosts of our disk transients, we subtract from our narrowband F665N images a scaled version of our F606W image. We note that the \Ha{} map will have also have some contribution from [N{\sc ii}] emission, however the contribution of these lines in star forming regions is small compared to the \Ha{} flux \citep{james05}. The scaling was determined by the ratio of the integrated filter transmission curves. Our observations are summarised in \cref{tab:observations}.

The ACS/WFC data were obtained from the archive\footnote{\url{https://archive.stsci.edu/hst/}} already drizzled and corrected for leakage due to charge transfer efficiency effects. We corrected the WFC3/UVIS exposures for this effect using {\sc wfc3uv\_ctereverse\_wSUB.F}\footnote{\url{http://www.stsci.edu/hst/wfc3/tools/cte\_tools}} and then drizzled them using {\sc DrizzlePac}\footnote{\url{http://drizzlepac.stsci.edu/}} to the native pixel scale of UVIS.

\begin{table}
\begin{threeparttable}
 \caption{\hst{} observations of \ca{} environments}
 \begin{tabular}{llll}
\hline
Transient & Instrument& Filter& Exp time (s) \\
\hline
SN~2001co & WFC3/UVIS & F438W & 990  \\
          &           & F606W & 990  \\ 
          &           & F665N & 2640 \\                            
SN~2003dg & WFC3/UVIS & F438W & 978  \\
          &           & F606W & 978  \\ 
          &           & F665N & 2625 \\
SN~2003dr & ACS/WFC   & F435W & 2460 \\
          &           & F606W & 2328 \\
SN~2005E  & ACS/WFC   & F435W & 2324 \\
          &           & F606W & 2192 \\
SN~2007ke & ACS/WFC   & F435W & 2416 \\
          &           & F435W & 2288 \\
\end{tabular}
\label{tab:observations}
\end{threeparttable}
\end{table}

\section{Data analysis}
\label{sect:analysis}
\subsection{Distances and reddening}

Distances were calculated using the recession velocity (\vrec{}) of the hosts\footnote{From the NASA/IPAC Extragalactic Database (NED), \url{http://ned.ipac.caltech.edu/}} and using $H_0 = 67.8$~\kms{}~Mpc$^{-1}$ and $\Omega_\text{m} = 0.308$ \citep{planck15}. We note that the typical uncertainties on the distance moduli of the hosts used ($\sim 0.1-0.2$~mag) do not significantly affect our conclusions based on photometry of the explosion sites.
For SN~2007ke, the host is uncertain. Two proximate early type galaxies, NGC~1129 and MCG+07-07-003, are the most probable candidates. Throughout the paper we adopt the distance of SN~2007ke to be that of NGC~1129, the slightly more distant (as inferred from recessional velocity) of the two.\footnote{The choice of host has little impact on our results for SN~2007ke. If using the distance of MCG+07-07-003, our absolute photometry would be $0.1$~mag deeper.} Galactic extinction was accounted for using a $R = 3.1$ Milky Way extinction law \citep{cardelli89} based on the dust maps of \citet{schlafly11}. For the remote sample, we assume negligible host galaxy extinction given their locations. The extinction of the disk sample is discussed in \cref{sect:diskenviron}.
Our adopted distances and reddening values are given in \cref{tab:distred}.

\begin{table}
\begin{threeparttable}
 \caption{Adopted distances and foreground reddening values for \cas{}}
 \begin{tabular}{lcccc}
\hline
Transient & \vrec{}  & $\mu$\tnote{a} & E(B-V)$_\text{MW}$\tnote{b} \\
          & (\kms{}) & (mag)          & (mag)                       \\
\hline
SN~2001co & 5166  & 34.44 & 0.017  \\
SN~2003dg & 5501  & 34.58 & 0.021  \\
SN~2003dr & 2237  & 32.61 & 0.013  \\
SN~2005E  & 2694  & 33.01 & 0.032  \\
SN~2007ke & 5194  & 34.45 & 0.095  \\
\end{tabular}
\label{tab:distred}
\begin{tablenotes}
 \item [a] Distance modulus
 \item [b] Galactic extinction from \citet{schlafly11}
 \end{tablenotes}
\end{threeparttable}
\end{table}

\subsection{Relative Astrometry}
\label{sect:astrometry}

In order to determine the location of the transients in our post-explosion imaging of the environments, we used offset sources provided by the discovery surveys, as well as astrometrically aligning imaging of the SNe, when available.

\subsubsection{SNe~2001co, 2003dg and 2003dr}

In order to determine the locations of these SNe in our \hst{} imaging, we used offsets from their respective host centres (SN~2001co: \citealt{iauc7643}, SN~2003dg: \citealt{iauc8113}, SN~2003dr: \citealt{iauc8117}). Considering the precision to which the offsets are given, and our findings for SNe~2005E and 2007ke below, we take an uncertainty of 0.2~arcsec in our determined locations for these SNe.

\subsubsection{SN~2005E}

To determine the location of SN~2005E in our \hst{} image we used imaging of the SN taken as part of the Carnegie Supernova Project \citep[CSP;][]{hamuy06}. An astrometic transformation between the CSP and \hst{} images was determined using 17 common sources as tie points. This resulted in an rms of 2.9~\hst{} pixels, or $\sim 0.15$~arcsec, in the transformation solution. The SN is well detected in the CSP image and as such the uncertainty from centroiding is very small compared to the transformation uncertainty. We take $0.15$~arcsec as our uncertainty in the location of SN~2005E in our \hst{} imaging. This was found to be consistent with the location determined using offsets for SN~2005E\footnote{\url{http://w.astro.berkeley.edu/bait/public_html/2005/sn2005E.html}}, provided by the same group whose offsets we used for SNe~2001co, 2003dg and 2003dr.

\subsubsection{SN~2007ke}
For SN~2007ke we used the offsets from NGC~1129 provided in \citet{chu07} in order to locate SN~2007ke in our \hst{} imaging. NGC~1129 is actually a double-cored early-type galaxy, with a separation of $\sim 0.8$~arcsec between the cores' peaks (see \cref{sect:morph}). We used the flux barycentre of the two peaks as the galaxy centre. To additionally confirm our astrometry, we used $r$-band SDSS images of the region in order to calculate the location based on offsets from a nearby star\footnote{\url{http://w.astro.berkeley.edu/bait/public_html/2007/sn2007ke.html}} (which was unfortunately not captured in our HST imaging), as well as repeating for the NGC~1129 offsets. Using 43 point sources we then made an accurate geometric transformation of the SDSS coordinates to our \hst{} image. We took the mean of the three locations in the \hst{} image and adopted this as our location of SN~2007ke. The uncertainty on the location was taken as the standard deviation of the individual determinations, which we found to be $0.15$~arcsec. We note that our results remain unchanged if we adopt any of the individual location determinations.

\subsection{Source detection and limiting magnitudes}

We utilised the \sex{} package \citep{bertin96} to perform source detection on our images. Detection was made on filtered images after applying a $3\times3$ FWHM gaussian. Detections were considered as islands of 5 or more pixels at $>1\sigma$ above the background level. 

In order to determine the limiting magnitude at the location of remote examples (\cref{sect:tranloc}), we used an aperture of radius 0.15~arcseconds. Aperture corrections were determined using bright, isolated point sources in the images to correct the magnitude limit to 0.5~arcsecond radius and then corrected to total magnitudes using enclosed energy fraction for ACS\footnote{\url{http://www.stsci.edu/hst/acs/analysis/apcorr}}. For the disk events we used apertures of 0.2~arcseconds radius and measurements were directly corrected to a large aperture using tabulated enclosed energy distributions\footnote{\url{http://www.stsci.edu/hst/wfc3/phot_zp_lbn}}. The sky background distribution was determined from a large annulus around the location of interest and we quote magnitude limits at the 3$\sigma$ level. This measurement was repeated using 1000 locations distributed around our assumed location based on our uncertainty on the location (\cref{sect:astrometry}). The spread of limiting magnitudes was minimal due to this shifting (a few hundredths of a mag in each case), meaning our results are unaffected by our location uncertainties for the remote sample. Magnitudes are given in the \hst{} VEGAmag system.

We also used {\sc DOLPHOT 2.0 ACS}\footnote{\url{http://americano.dolphinsim.com/dolphot/}} in order to detect objects in our imaging, with standard parameters for the ACS module. The results from {\sc DOLPHOT} also gave another estimate of the magnitude limits based on recovery of fake star insertions in the images, which agreed well with our estimates from using a local background estimate. 

We note here the expected morphology of our sources of interest at the explosion sites. For typical GC radii of a few--ten pc, these will appear as point sources in our \hst{} images. Dwarf satellite galaxies, however, can span a range of effective radii. Even for a conservative radius of 100~pc, more typical of very faint dwarf satellites, we would resolve such sources even for the most distant of our sample.

\section{Results}
\label{sect:results}

\subsection{Limits and nearby detections at the locations of remote \cas{}}
\label{sect:tranloc}

\begin{figure*}
 \centering
 \includegraphics[width=0.33\textwidth]{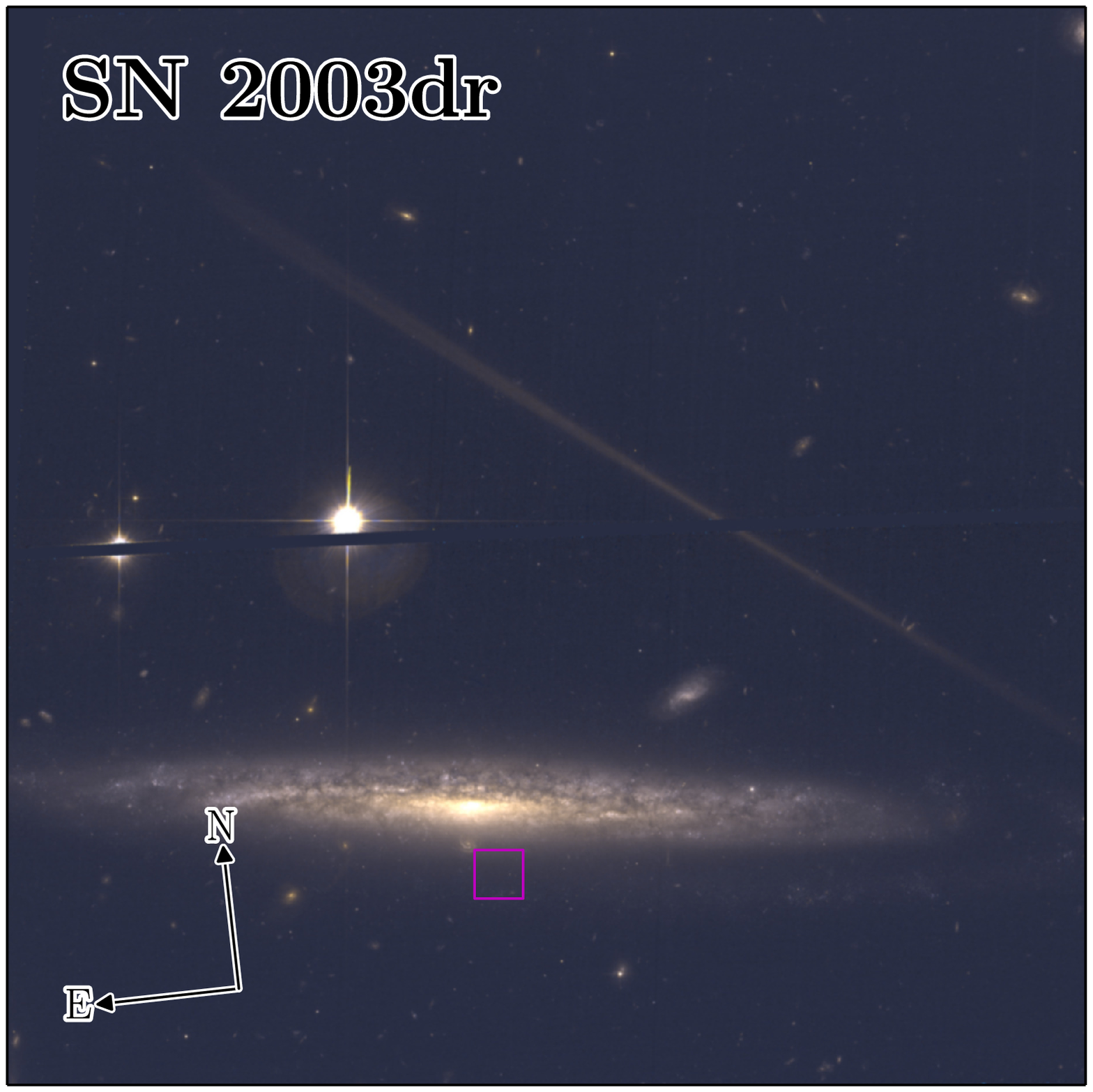}
 \includegraphics[width=0.33\textwidth]{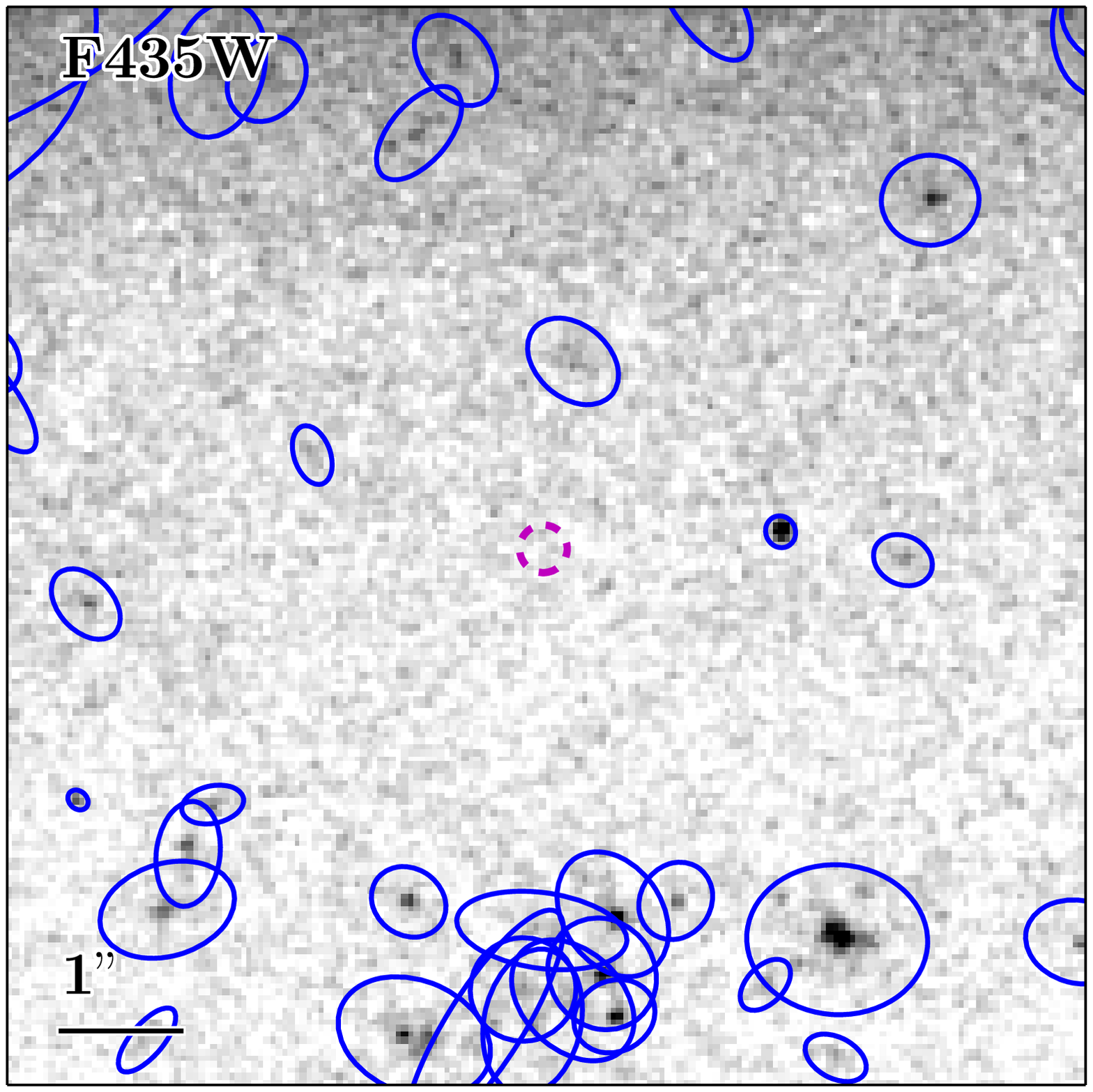}
 \includegraphics[width=0.33\textwidth]{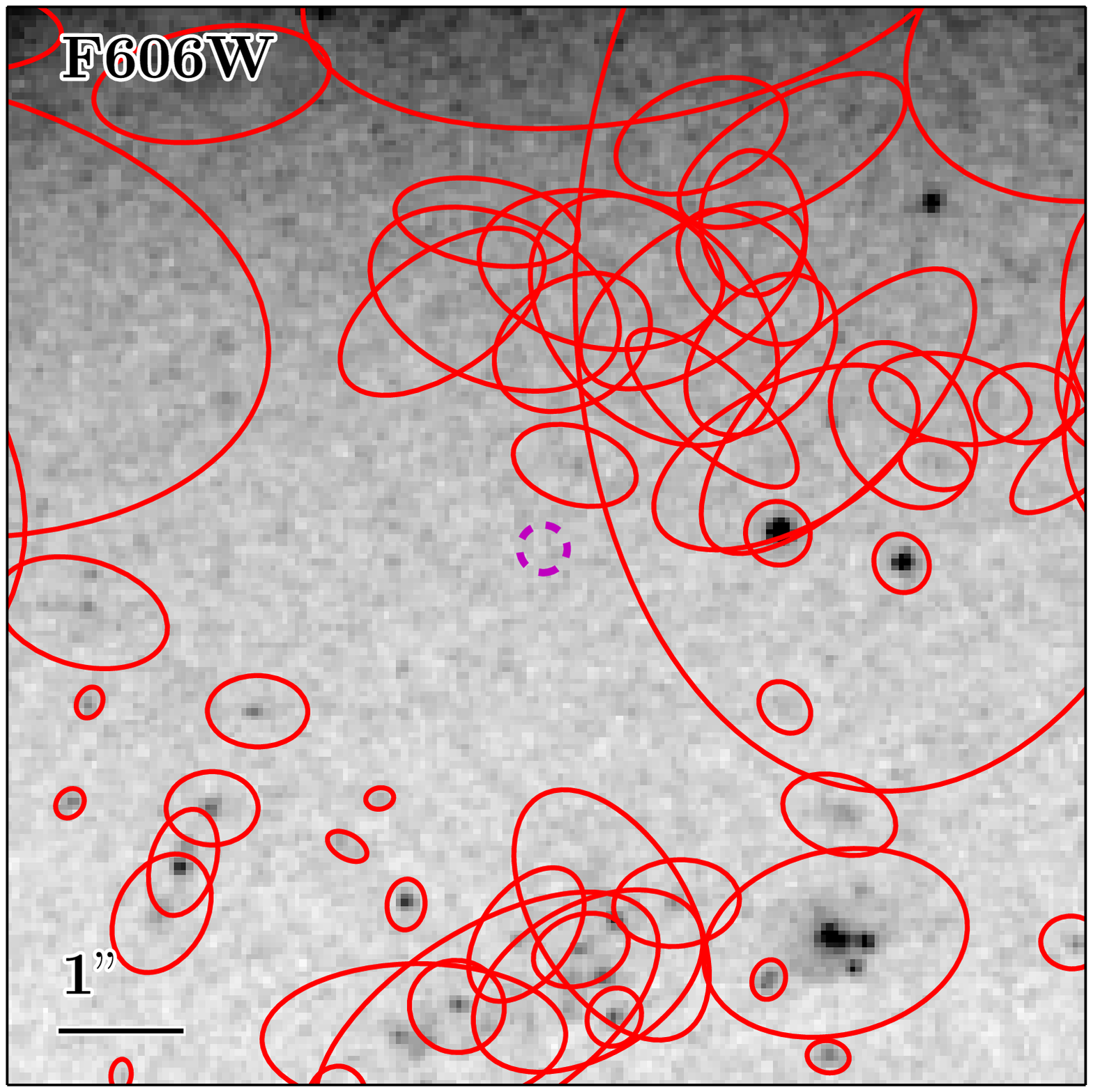}
 \caption{\hst{} imaging (blue: F435W, green: \texttt{average}(F435W, F606W), red: F606W) of the host, NGC~5714, and immediate environment of SN~2003dr. The host image is 200~arcsec on a side, and the sizes of the individual filter images are indicated by the magenta square. The location of SN~2003dr is shown at the centre of each zoomed image, with the dashed circle having a radius of $0.2$~arcsec. Kron apertures of \sex{} detected sources are marked on their respective images, which have been slightly smoothed to aid visual identification of sources. At the distance of NGC~5714 one arcsec is $\sim 160$~pc.}
 \label{fig:2003dr_loc}
\end{figure*}

\begin{figure*}
 \centering
 \includegraphics[width=0.33\textwidth]{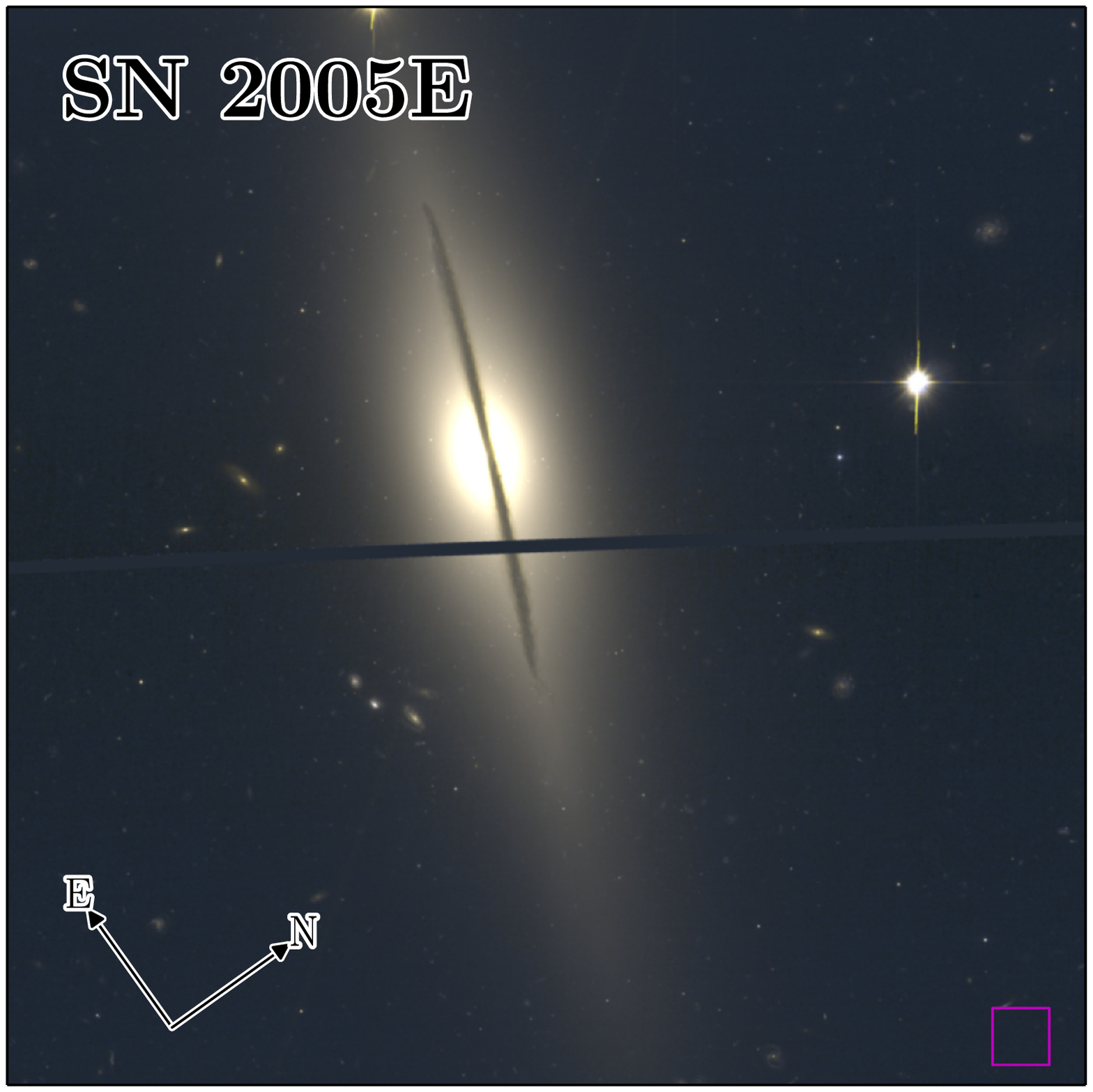}
 \includegraphics[width=0.33\textwidth]{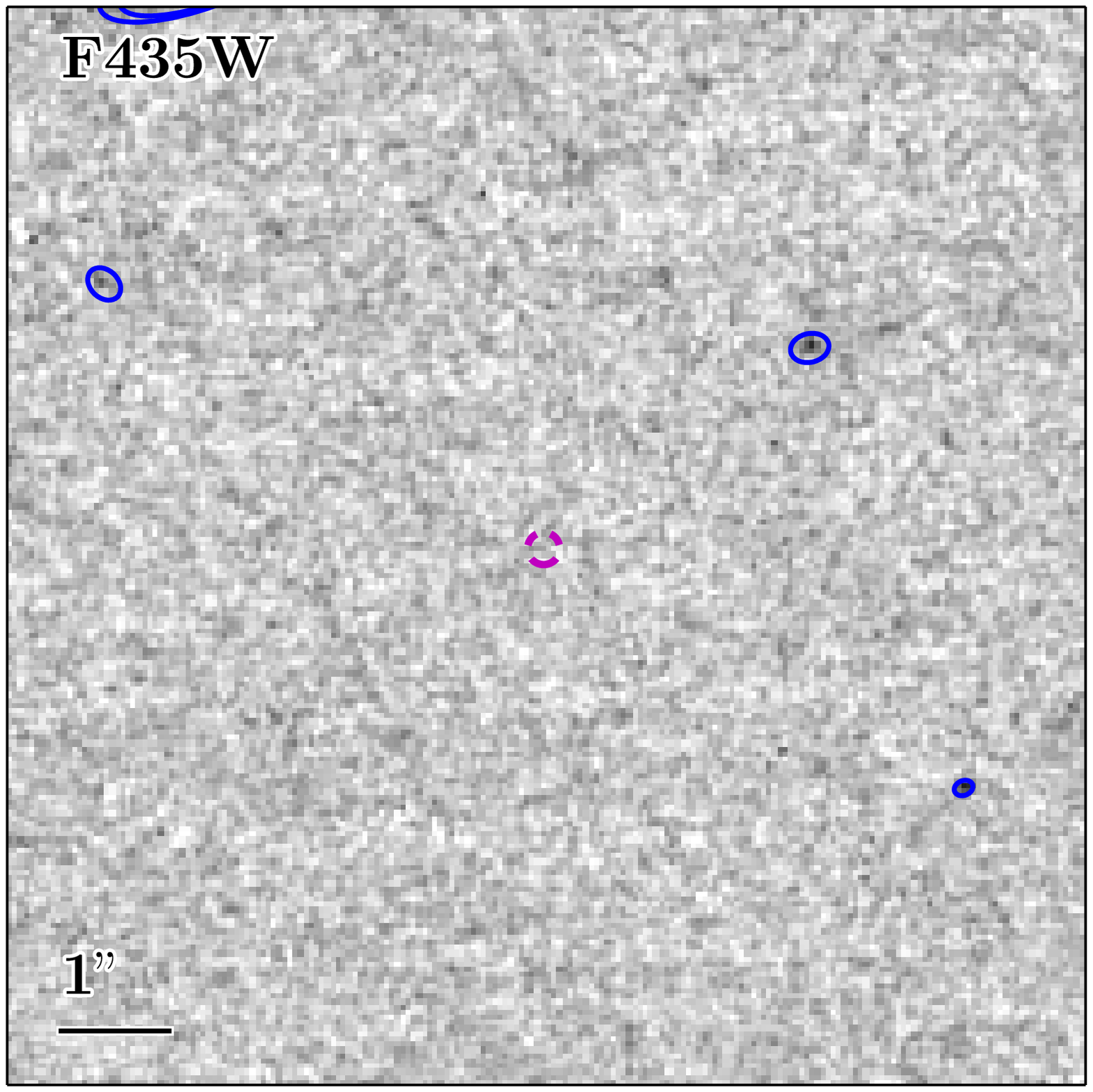}
 \includegraphics[width=0.33\textwidth]{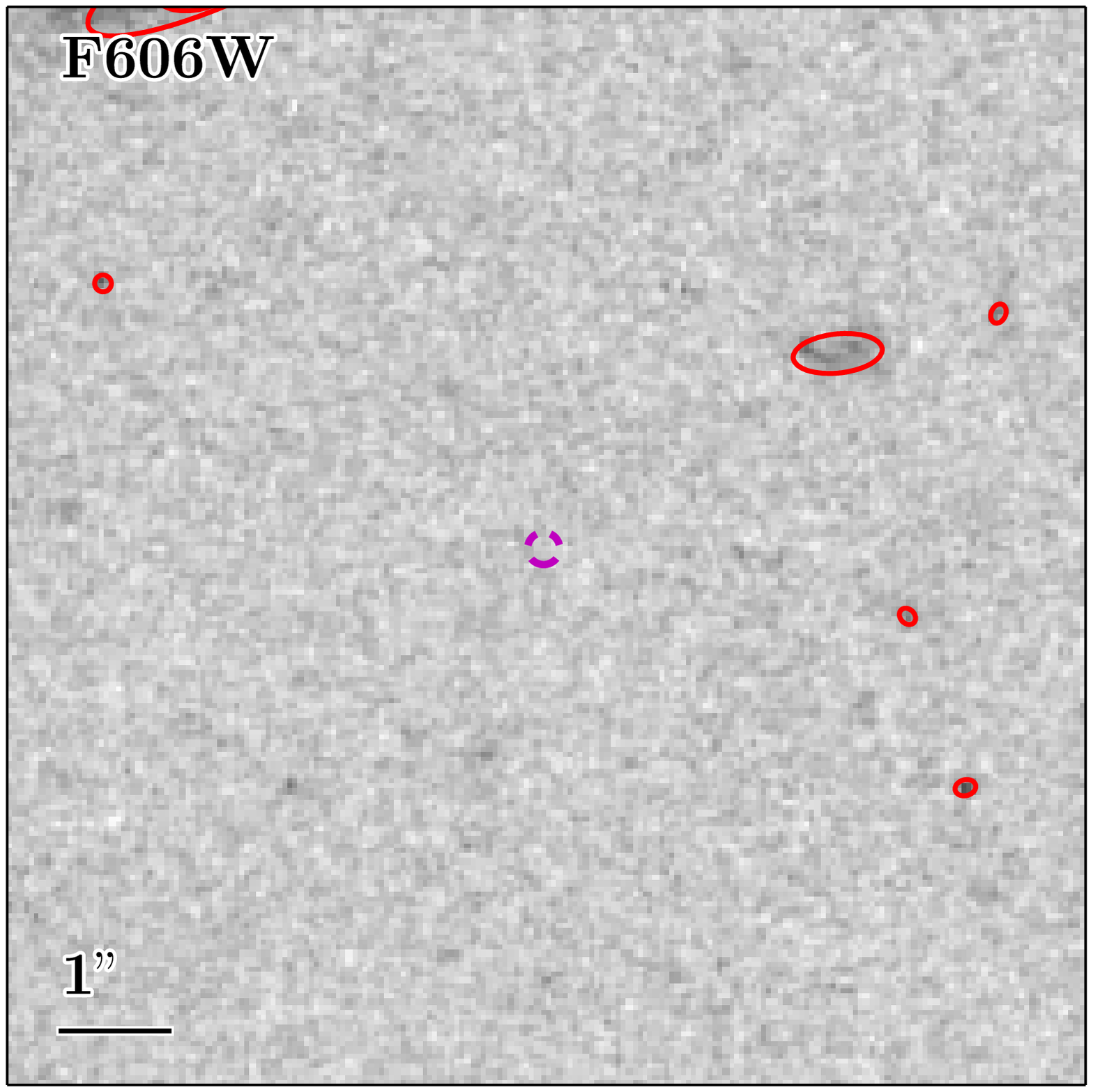}
 \caption{As \cref{fig:2003dr_loc} for SN~2005E. The host image is 190~arcsec on a side and the dashed circle marking the location of SN~2005E has a radius of 0.15~arcsec, our estimate of the uncertainty. At the distance of NGC~1032 one arcsec is $\sim 190$~pc.}
 \label{fig:2005E_loc}
\end{figure*}

\begin{figure*}
 \centering
 \includegraphics[width=0.33\textwidth]{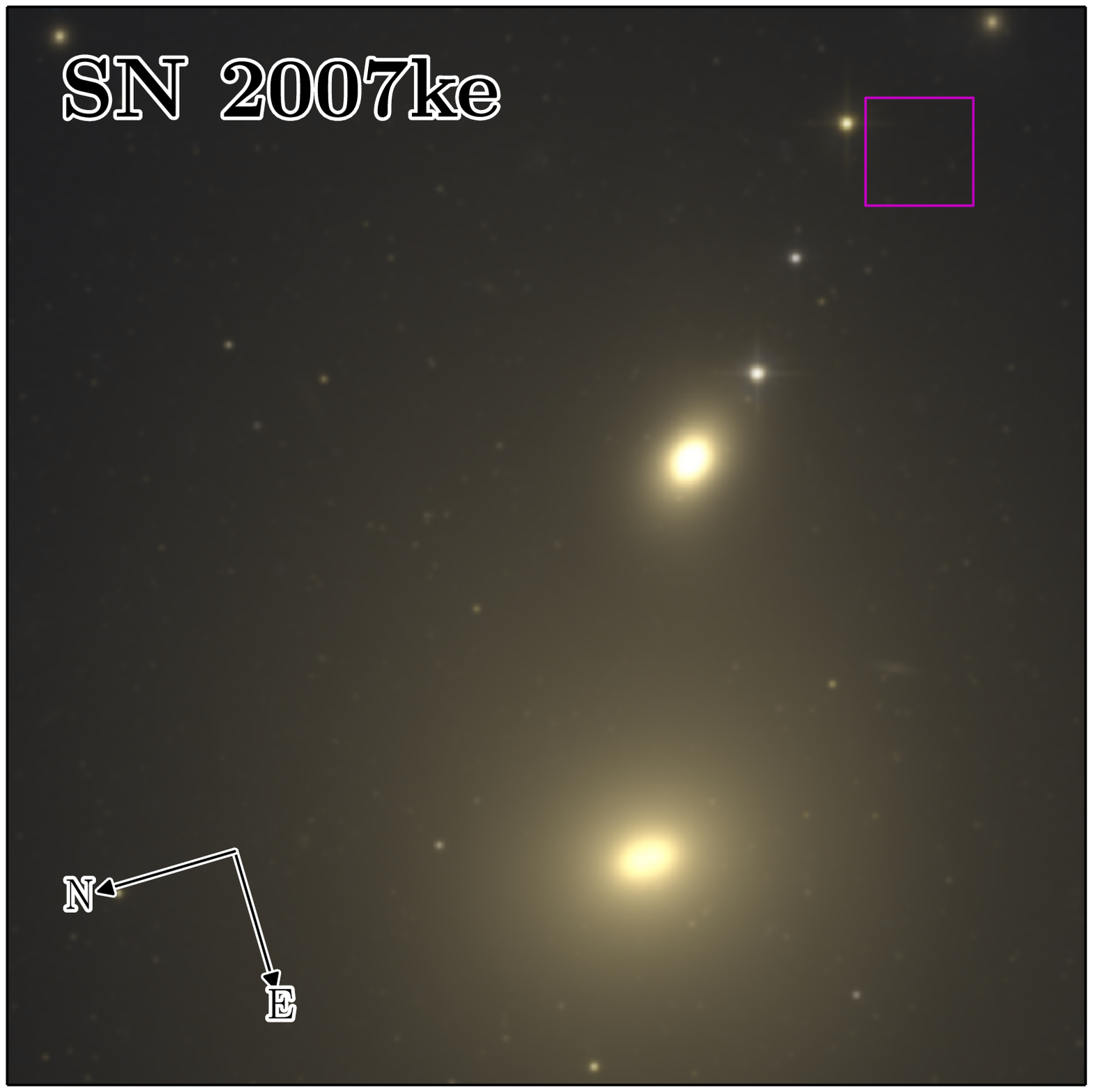}
 \includegraphics[width=0.33\textwidth]{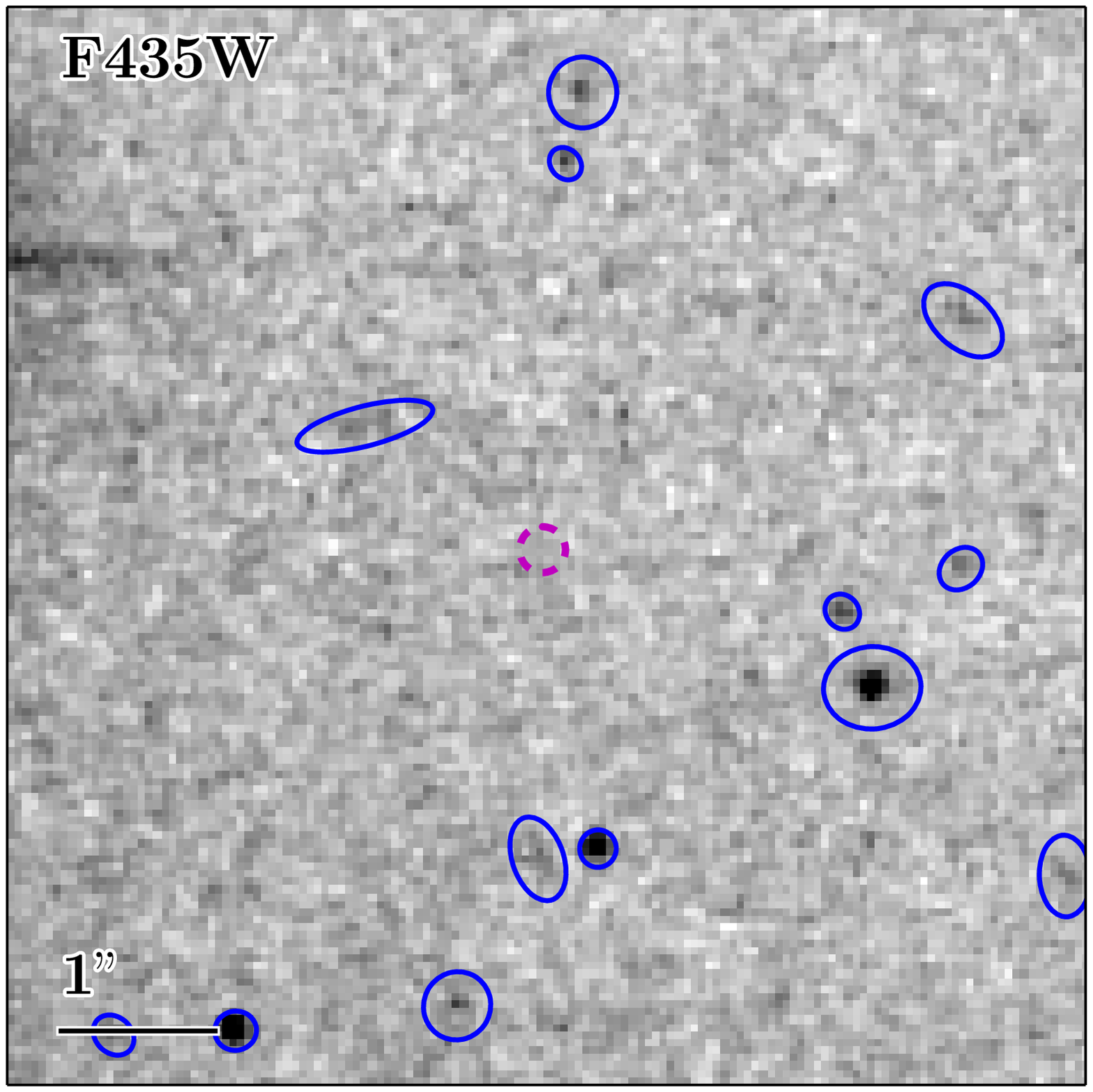}
 \includegraphics[width=0.33\textwidth]{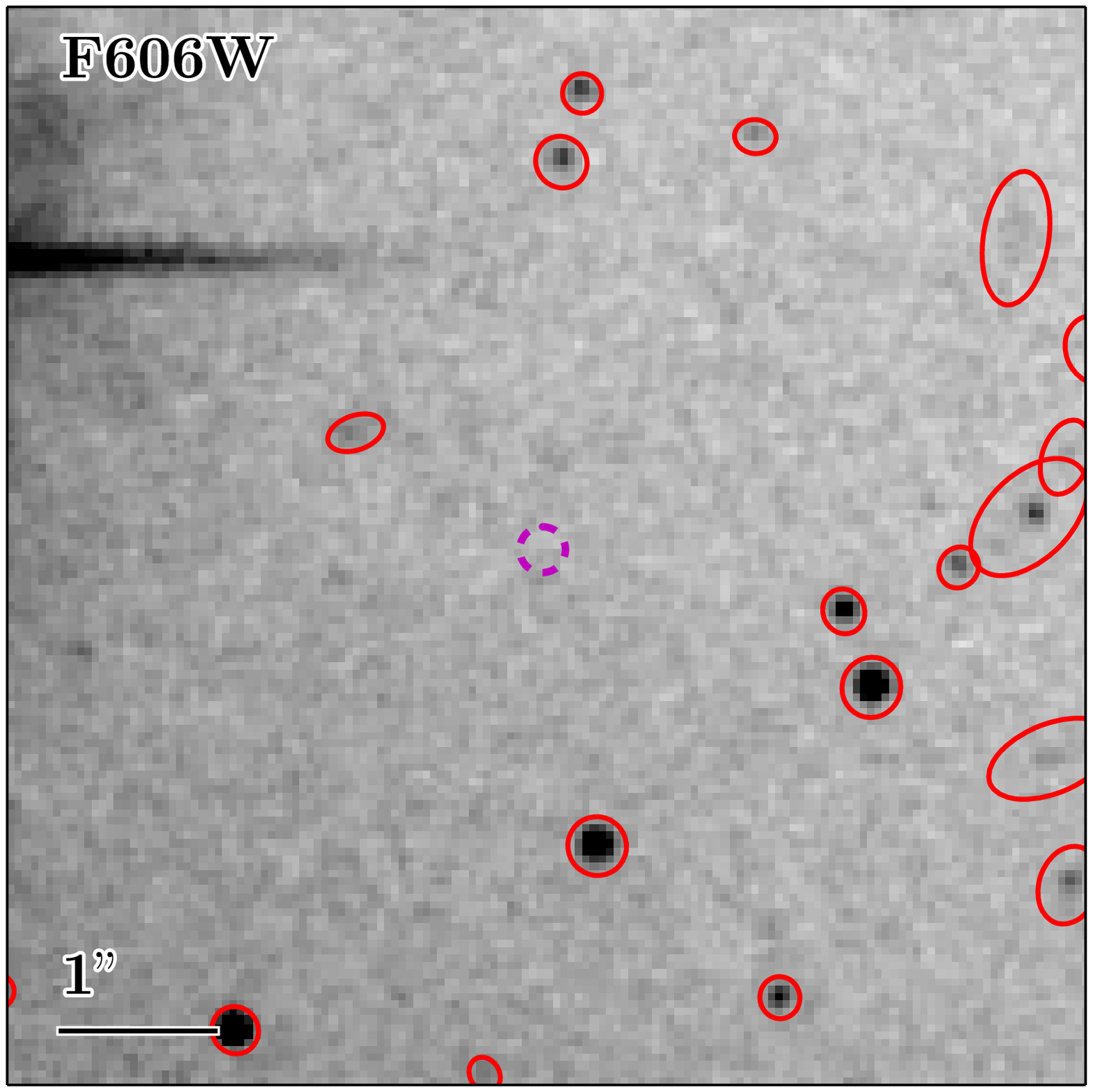}
 \caption{As \cref{fig:2005E_loc} for SN~2007ke. The host image is 70~arcsec on a side and the dashed circle marking the location of SN~2007ke has a radius of 0.15~arcsec, our estimate of the uncertainty. At the distance of NGC~1129 one arcsec is $\sim 380$~pc.}
 \label{fig:2007ke_loc}
\end{figure*}

We did not detect an underlying source in any of our images for SNe~2003dr, 2005E or 2007ke. Our limits are presented in \cref{tab:maglims} and images, showing local \sex{} detected sources, are plotted in \cref{fig:2003dr_loc,fig:2005E_loc,fig:2007ke_loc} for SNe~2003dr, 2005E and 2007ke, respectively.

The environment of SN~2003dr is quite complex. Although relatively close in linear distance or isophotal radius terms, SN~2003dr lies offset along the minor axis of the galaxy and thus off the disk light. Our \hst{} imaging shows a strong tidal feature that passes through the location of the transient along the southern and western sides of NGC~5714. This feature is discussed in \cref{sect:morph}. In particular for F606W, the \sex{} detection routine assigns many distinct regions of the host galaxy's underlying light as separate objects. Nevertheless, even with this somewhat aggressive detection and deblending, there is no object found underlying the transient's location. The nearest detection is in the F606W image, 0.75~arcsec away, which is $\sim$~120~pc at the distance of NGC~5714. Thus, although there is evidence of a tidal feature coincident with the explosion site of SN~2003dr, we find no evidence of a compact cluster or dwarf hosting underlying

For SN~2005E, we presented a limit of $m_R > 27.4$~mag in \citet{lyman14d}. Even with increased depth and the addition of the bluer F435W band, the nearest detection remains the object north and slightly east of SN~2005E at a distance of $\sim$3.5~arcsec (which is extended in our F606W imaging). Our new photometry improves on the previous limit for SN~2005E \citep{lyman14d} and also provides a limit in the bluer F435W filter.

Although observations of SN~2007ke were taken with the same setup as SN~2005E and SN~2003dr, the limits are shallower owing to the extended halo of the nearby galaxies that is present at the transient's location and the higher degree of Galactic extinction towards SN~2007ke. The nearest detection appears in both F435W and F606W as an extended object, and is $\sim$1.5~arcsec, or $\sim$570~pc in plane-of-sky at the distance of NGC~1129.

\begin{table}
\begin{threeparttable}
 \caption{Point source magnitude limits at the locations of \cas{}\tnote{a}}
 \begin{tabular}{lcccc}
\hline
Transient & $m_{\rm F435W}$ & $M_{\rm F435W}$ & $m_{\rm F606W}$ & $M_{\rm F606W}$ \\
          & (mag)           & (mag)           & (mag)           & (mag)           \\
\hline
SN~2001co & $>26.5$        & $>-7.9$        & $>26.8$        & $>-7.6$        \\
SN~2003dg & $>26.2$        & $>-8.4$        & $>25.9$        & $>-8.7$        \\
SN~2003dr & $>28.1$        & $>-4.5$        & $>28.2$        & $>-4.4$        \\
SN~2005E  & $>28.0$        & $>-5.0$        & $>28.4$        & $>-4.6$        \\
SN~2007ke & $>27.7$        & $>-6.7$        & $>27.8$        & $>-6.6$        \\
\end{tabular}
\label{tab:maglims}
\begin{tablenotes}
 \item [a] Corrected for Galactic extinction from \citet{schlafly11}, see \cref{tab:distred}
 \end{tablenotes}
\end{threeparttable}
\end{table}

\subsection{The disk environments of SNe~2001co and 2003dg}
\label{sect:diskenviron}
\begin{figure*}
 \begin{minipage}{0.5\textwidth}
 \includegraphics[width=\linewidth]{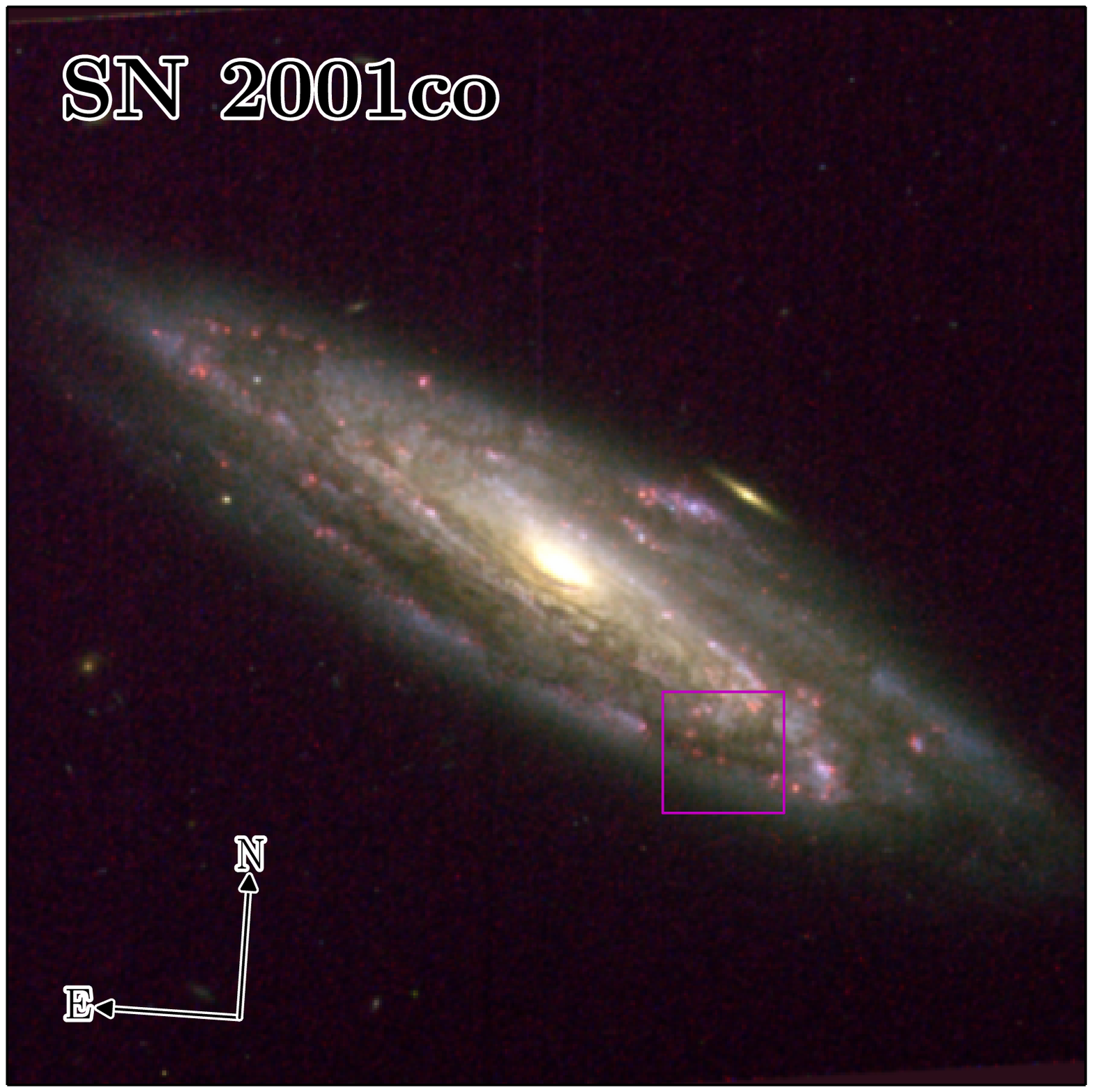}
 \end{minipage}%
 \begin{minipage}{0.5\textwidth}
 \includegraphics[width=0.49\linewidth]{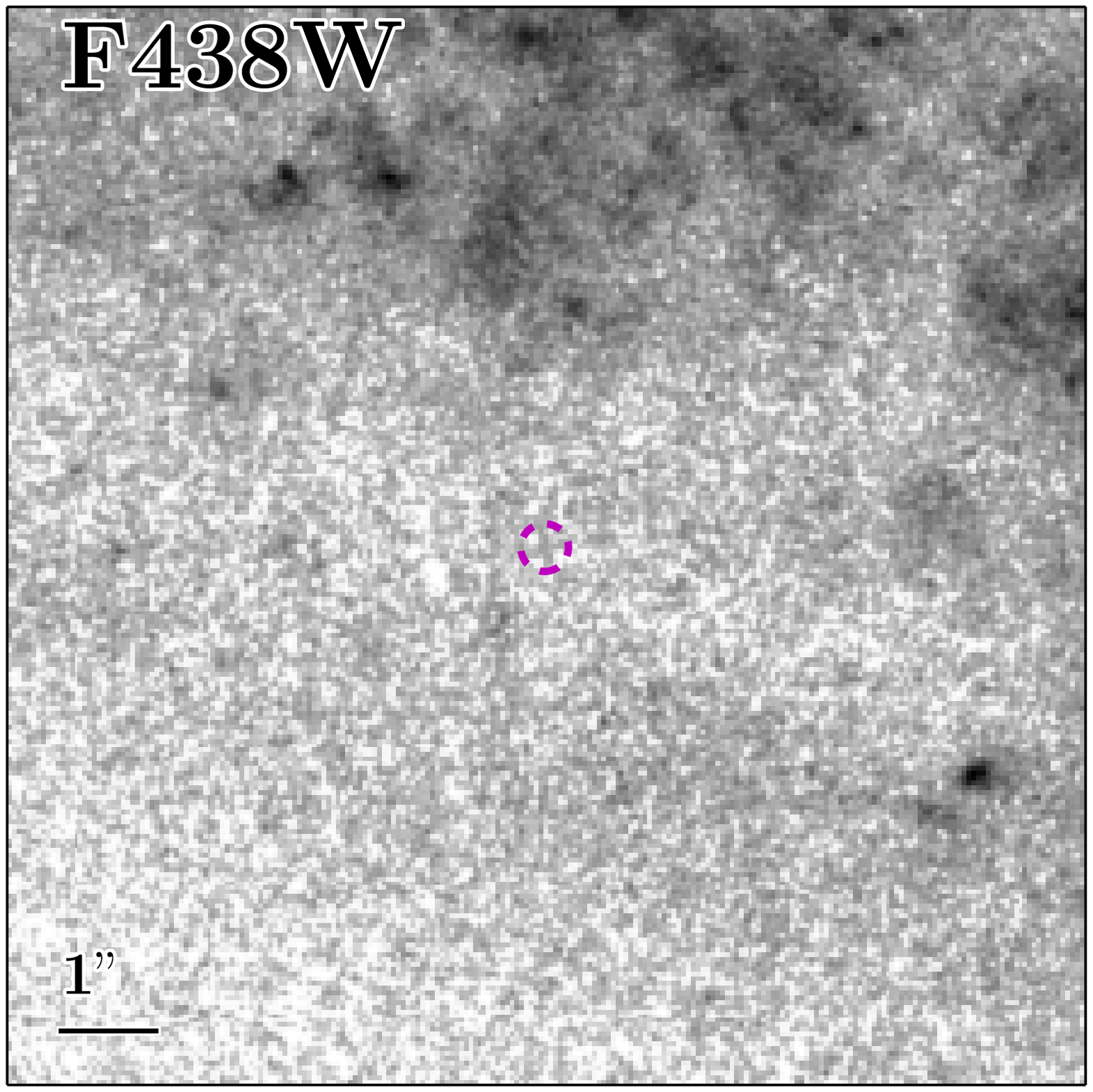}
 \includegraphics[width=0.49\linewidth]{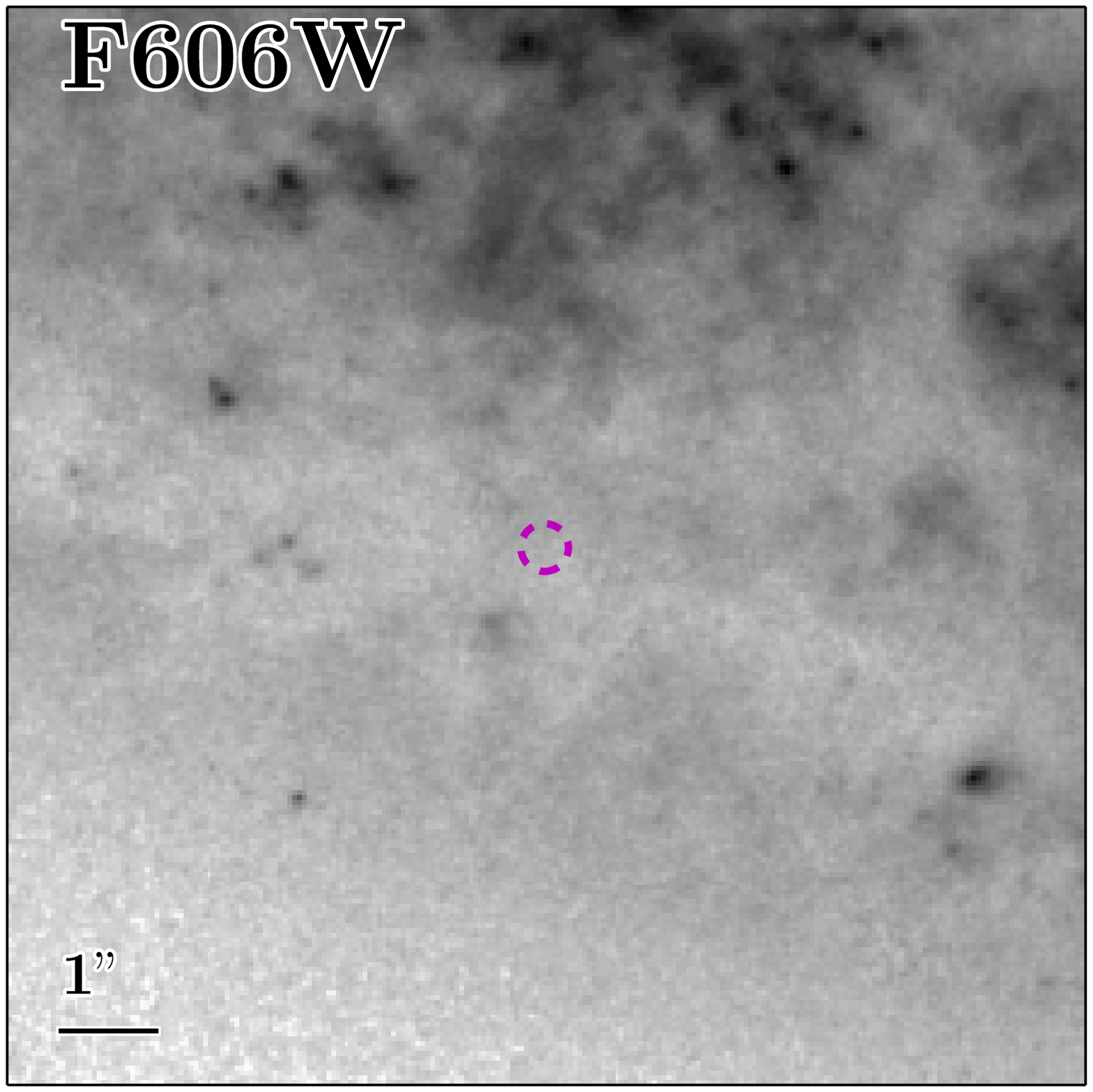}\\
 \includegraphics[width=0.49\linewidth]{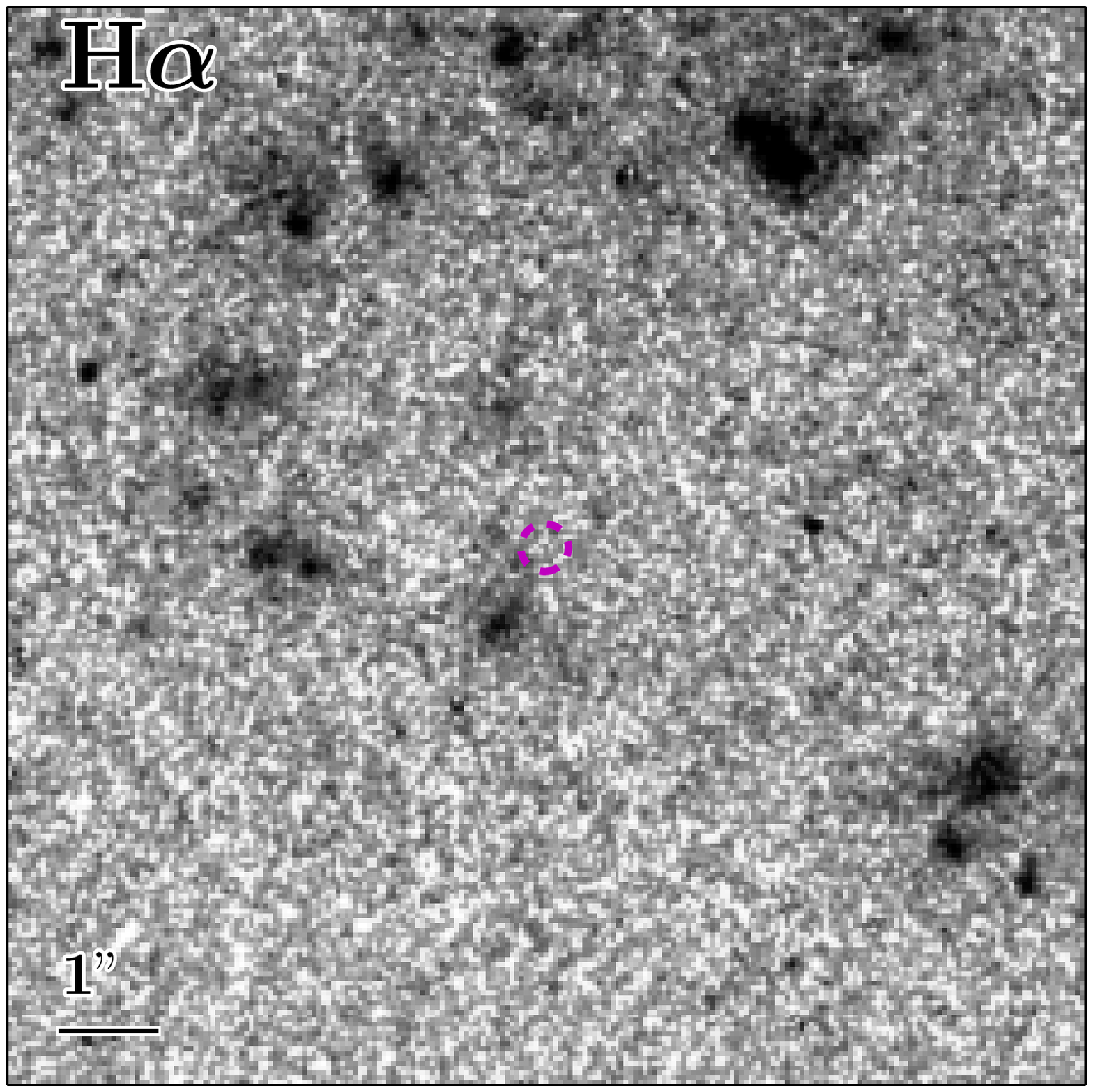}
 \end{minipage}
 \caption{\hst{} imaging (blue: F438W, green: F606W, red: F665N) of the host, NGC~5559, and immediate environment of SN~2003dr. The host image is 80~arcsec on a side, and sizes of the zoomed images are indicated by the magenta square. The \Ha{} map was made through the subtraction of a scaled version of F606W from the narrowband F665N image (\cref{sect:obs}). The location of SN~2001co is indicated at the centre of each zoomed image, with the dashed circle having a radius of $0.2$~arcsec. The zoomed images are slightly smoothed to aid visual identification of sources. At the distance of NGC~5559 one arcsec is $\sim 370$~pc.}
 \label{fig:2001co_loc}
\end{figure*}

\begin{figure*}
 \begin{minipage}{0.5\textwidth}
 \includegraphics[width=\linewidth]{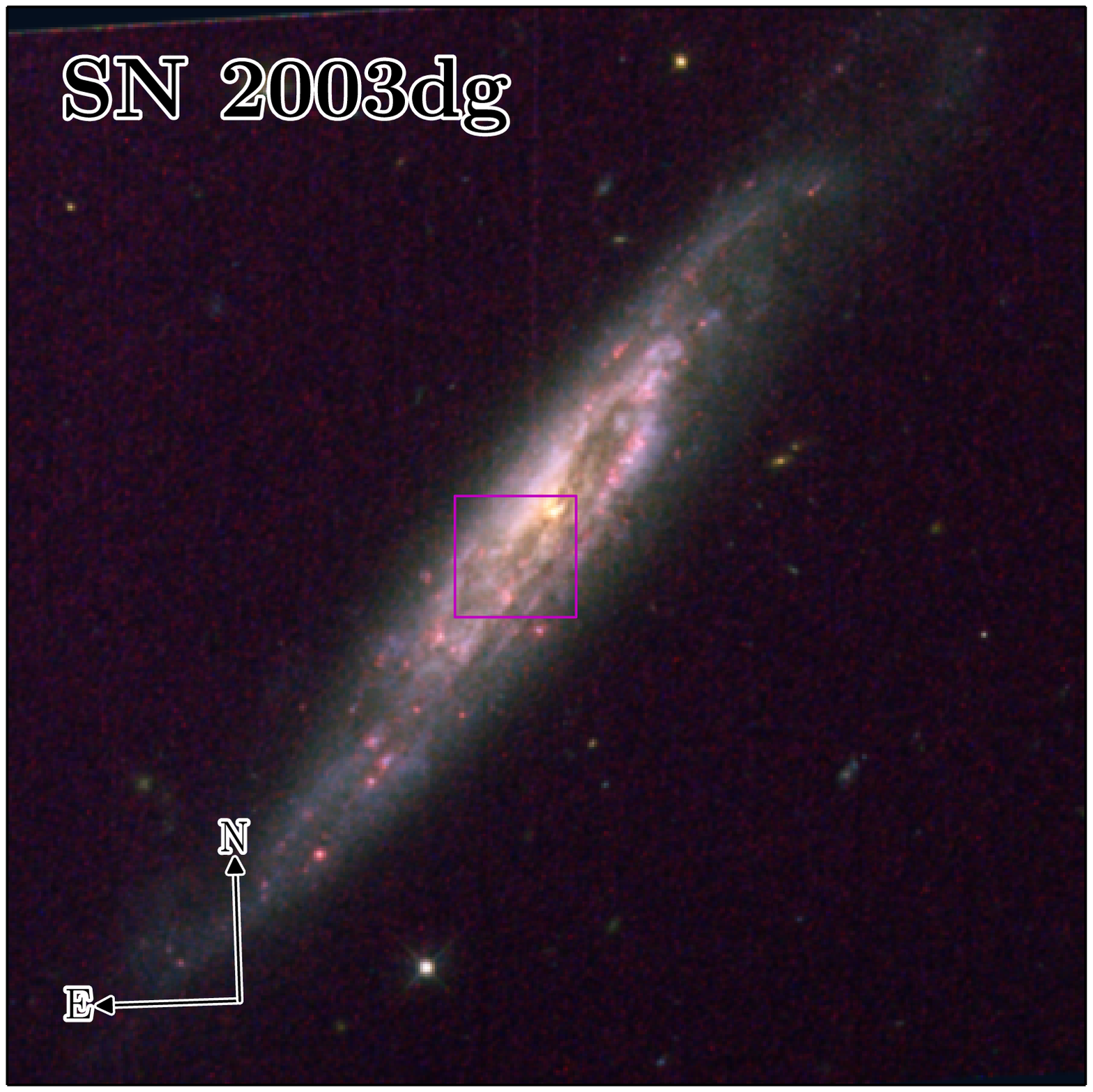}
 \end{minipage}%
 \begin{minipage}{0.5\textwidth}
 \includegraphics[width=0.49\linewidth]{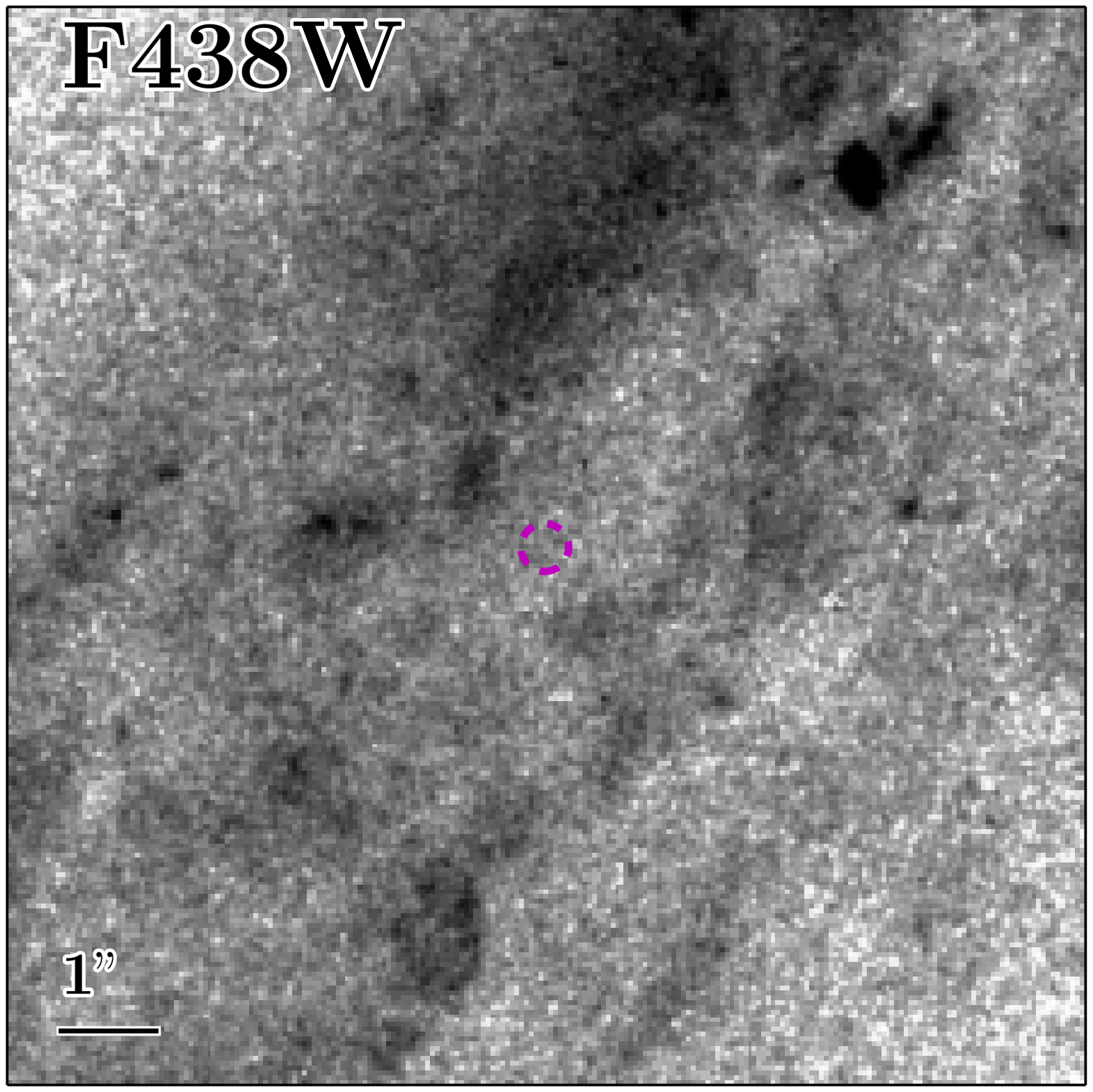}
 \includegraphics[width=0.49\linewidth]{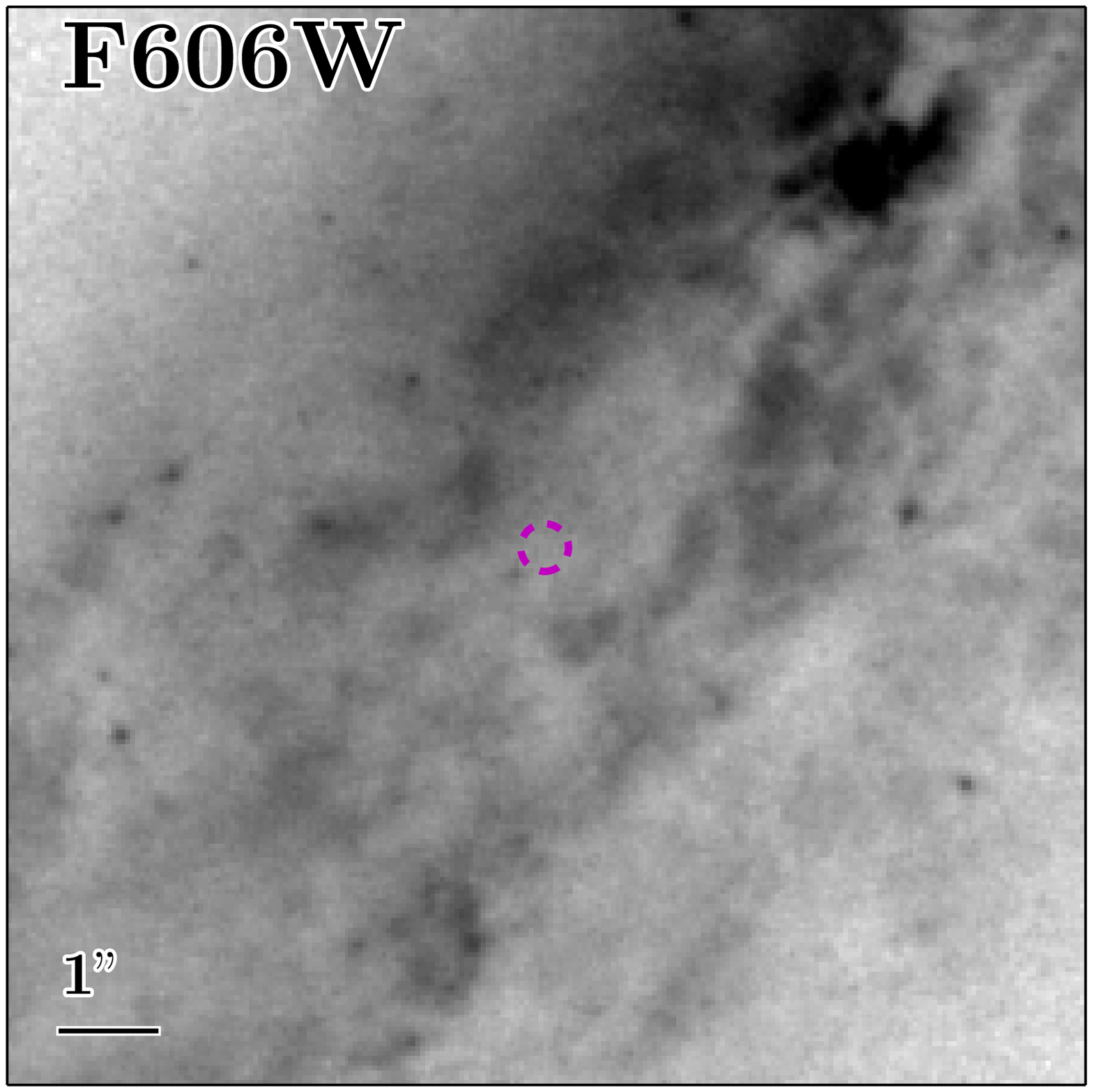}\\
 \includegraphics[width=0.49\linewidth]{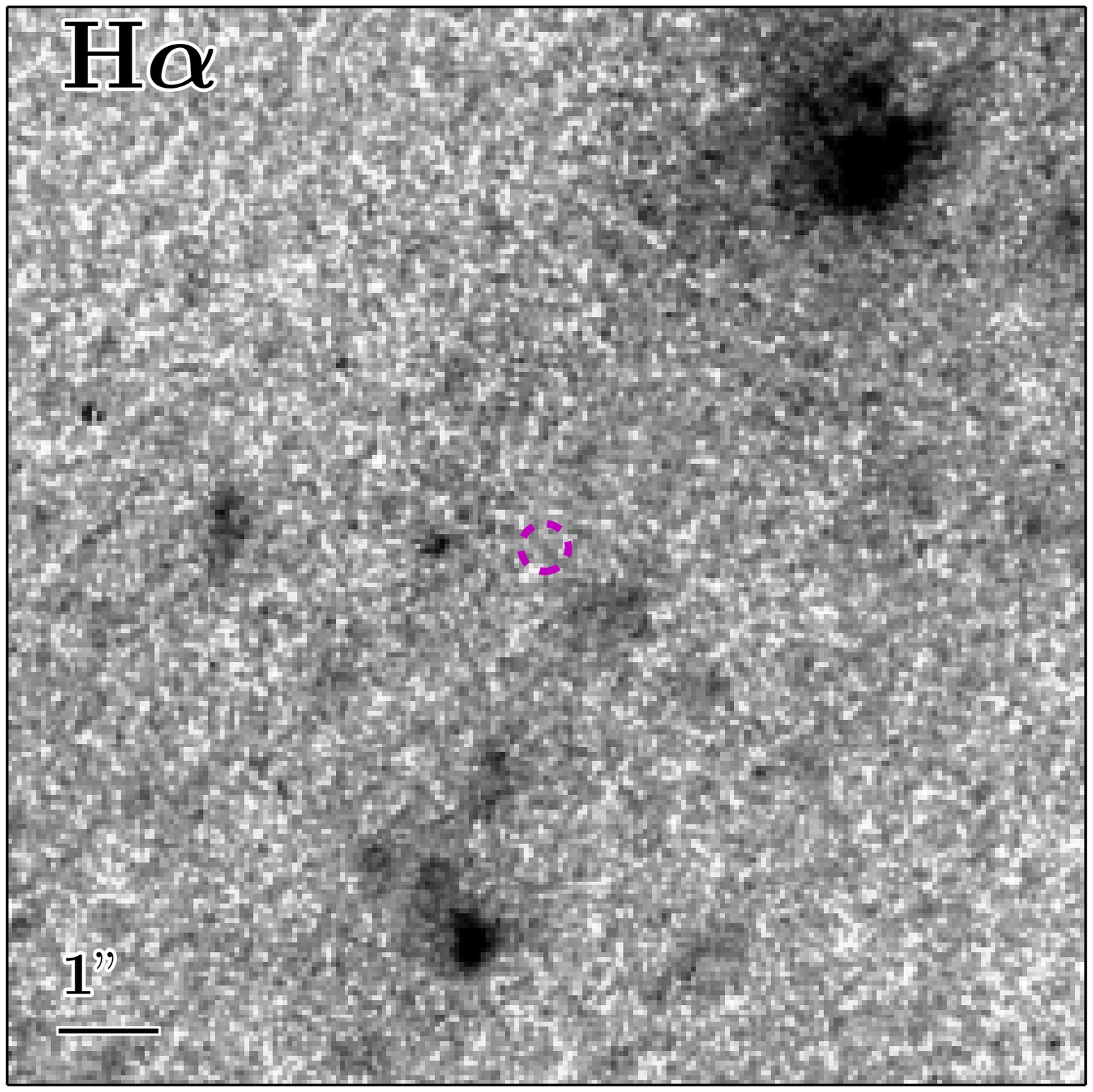}
 \end{minipage}
 \caption{As \cref{fig:2001co_loc} for SN~2003dg. At the distance of UGC~6934 one arcsec is $\sim 400$~pc.}
 \label{fig:2003dg_loc}
\end{figure*}

Two of the smallest offset examples of the Ca-rich class are SNe~2001co and 2003dg. Both lie within the disks of their hosts, which look rather similar -- slightly inclined, disk-dominated late type galaxies. Both display strong patchy star formation and significant dust lanes, as is typical for their morphological types. The hosts and nearby environments of these examples are shown in \cref{fig:2001co_loc,fig:2003dg_loc}.

The explosion site of SN~2001co is coincident with a strong dust lane. This attenuation results in a faint background at the explosion site, particularly in the bluer F438W filter. There is no distinguishable source underlying the location of SN~2001co in either F438W or F606W. The \Ha{} map shows nearby regions of strong star formation, in particular a region $\sim$270~pc to the south east, but nothing underlying the explosion site.

Similarly, the location of SN~2003dg is also in line-of-sight alignment with a central dust lane of its host, UGC~6934. Again there appears to be no evidence of an underlying source in the continuum bands or \Ha{} map at the explosion site.

If we assume no host galaxy extinction, as was done for the remote sample, then we can place limits on the presence of underlying compact sources, although these will be weaker given the relatively brighter background
and our shorter exposures (cf. the remote sample). These point source limits are presented in \cref{tab:maglims} for F438W and F606W bands for each event. These limits, although shallower, still preclude the presence of brighter underlying clusters. However, these limits refer only to a compact underlying source since, unlike in the case of the remote events, there is also clearly an underlying stellar population. Using our continuum-subtracted narrow-band image, we place a limit on emission flux over a 0.2~arcsecond aperture. This gives us a limit on an underlying emission region to be $L_{\text{H}\alpha} < 10^{35}$~erg~s$^{-1}$.

Although we take the extreme case of no host galaxy extinction, despite there locations being coincident with dust lanes, there are indications that the SNe did not suffer very large extinction. Their observed peak $B$-band magnitudes of $-15.09$ and $-15.03$ (for SNe~2001co and 2003dg, respectively), lie within the range of other \cas{}, $-14.04$ to $-15.45$~mag \citep{perets10}. Further, the kinematic study of \citet{foley15} found these events were both blueshifted compared to their hosts, indicating they are on the `near side' of the galaxy, limiting the amount of intervening dust. Thus, although these limits are likely to be intrinsically shallower for the true SN explosion sites, owing to some amount of unaccounted for extinction, they remain indicative values since we do not expect huge amounts of reddening.

\subsection{Host morphologies}
\label{sect:morph}
The deep and high-resolution images of \ca{} hosts has revealed some interesting morphological features, which were not detected or were ambiguous in previous imaging. 

\begin{figure}
 \centering
 \includegraphics[width=\columnwidth]{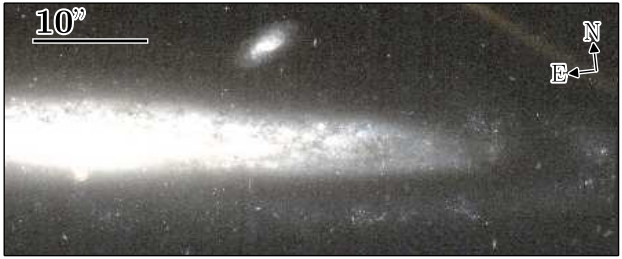}
 \caption{An image of NGC~5714 (blue: F435W, green: \texttt{average}(F435W, F606W), red: F606W), the host of SN~2003dr, showing the tidal feature to the south and west of the main disk. Note that the streak in the upper right is a instrumental artefact.}
 \label{fig:2003dr_tidal}
\end{figure}

NGC~5714, the host of SN~2003dr, appears relatively isolated from other large galaxies -- a group of galaxies is located $\sim$5~arcmin east, however these galaxies are more distant (\vrec{}$\sim$10000~\kms{} cf. for NGC~5714 \vrec{}~$= 2237$~\kms{}). Nevertheless there is a clear tidal feature distinctly separate from the main disk and offset to the south and west. Given the lack of nearby gravitational influences capable of causing such disruption to NGC~5714, this is most likely the remnant stream due to a minor merger or complete tidal disruption of a smaller satellite galaxy.\footnote{Although there is an apparent satellite dwarf galaxy to the north-west, it is well-formed and of insufficient mass to affect the disk of NGC~5714 significantly, discrediting it as the cause of the tidal feature.} The stream is shown in \cref{fig:2003dr_tidal} and appears to show strong, patchy star formation along its length. The stream is just detected in SDSS imaging, which also reveals a faint extended stream protruding from the east of the galaxy -- perhaps a continuation of the same feature seen on the south and west edges. Unfortunately the limited field of view of our \hst{} imaging meant we have not captured this eastern feature. We have inspected a CALIFA \citep{sanchez12} data cube of NGC~5714, which was initially analysed as part of a statistical study by \citet{galbany14}. These data indeed confirm regions of intense star formation in the tidal stream as revealed by strong \Ha{} emission, most prominently from the region to the south of the SN location (\cref{fig:2003dr_loc}). The wavelength of the \Ha{} peak emitted from this region matches that of the \Ha{} peak of the galaxy centre, limiting the velocity offset between the two to being very small ($<$150~\kms{}, \citealt{sanchez12}). The velocity offset of SN~2003dr from NGC~5714, based on the profile of the [Ca{\sc ii}] $\lambda$7291, 7324 emission, is $560\pm40$~\kms{} \citep{foley15}. Although this value is subject to significant and difficult to quantify uncertainty arising from, e.g., the ejecta distribution of SN~2003dr and the shape of the underlying continuum, it is indicative that the SN is not kinematically related to the tidal feature, and potentially offset in the line-of-sight direction. Alongside the lack of an underlying source detected in our \hst{} imaging, this is further evidence of the remote nature of this example of the class. A lack of underlying star formation at the explosion site also agrees with the match of an elliptical template spectrum to the nebular continuum of SN~2003dg that was found by \citet{foley15}.

There also appear to be extended features along the major axis of UGC~6934 (the host of SN~2003dg), particularly on the north west side. However, this is a fairly minor morphological feature and the main disk appears well formed.

\begin{figure}
 \centering
 \includegraphics[width=0.75\columnwidth]{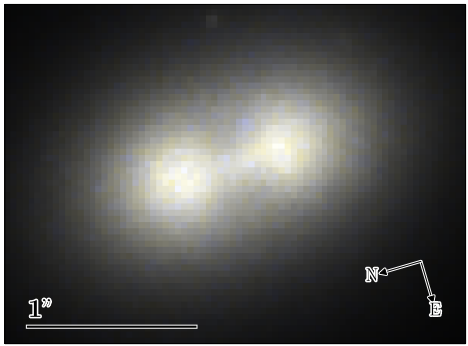}
 \caption{A false colour image (blue: F435W, green: \texttt{average}(F435W, F606W), red: F606W) zoomed in on the double core of NGC~1129, the presumed host of SN~2007ke. Scaling has been stretched to highlight the morphology of the bright cores.}
 \label{fig:2007ke_doublecor}
\end{figure}

The centre of NGC~1129, the host of SN~2007ke, is clearly resolved into a double core in \hst{} imaging, as shown in \cref{fig:2007ke_doublecor}. The disturbed nature of NGC~1129 was previously noted by \citet{peletier90}. These authors found a large (90 degrees) change in the positional angle of its light profile with radius, which they suggested was the result of a recent merger -- the extreme positional angle change prompting them to speculate that NGC~1129 could be the result of a major merger of two comparable ellipticals. Indeed, based on simple measurements of the peak brightness, aperture flux and F435W--F606W colour of the individual cores, we find them to be extremely similar, in support of this early suggestion. The cores are separated by $\sim$0.85~arcsec, or 320~pc at the distance of NGC~1129, as shown in \cref{fig:2007ke_doublecor}.

Such morphological discussion of the hosts is pertinent in light of the recent study of \citet{foley15}. \citeauthor{foley15} noted the high fraction of recent merger activity amongst the hosts of \cas{}, as well as an over-abundance of AGN compared to field galaxies. A causal link was posited between the rate of \cas{} in those galaxies showing these features. We have shown here NGC~1129 is an ongoing (likely to be major) merger through its double-cored nature, and NGC~5714 harbours clear signs of a recent minor merging -- thereby increasing the fraction of \ca{} hosts with firm signs of ongoing merging activity. 

\section{Discussion}
\label{sect:discuss}

The study of \citet{lyman14d} presented limits on potential underlying host systems for nearby \cas{}. These results ruled out a number of previous progenitor suggestions and explanations for their extreme offset distribution \citep[e.g.][]{kasliwal12,yuan13} for those events. Our results here build on these findings through a lack of detections at the explosion sites, providing two additional examples with deep limits (SNe~2003dr and 2007ke), plus deeper limits in additional bands for SN~2005E, thereby producing a more complete census of the environments of the remote members of this class. Additionally, we have investigated two examples of the class located coincident with the disk of their hosts. Combined with literature results \citep{maund05,perets11,kasliwal12,lyman14d}, there has yet to be a detection of an obvious birth site (e.g. star formation region, GC or satellite galaxy) directly underlying the location of any \ca{} explosion site that has been investigated. Six of these have limits deep enough to rule out the presence of various host systems for those events individually and, when combined, the \ca{} class as a whole.

Following \citet{lyman14d}, the new limits presented here give further evidence against the now largely defunct hypothesis of massive stars as the progenitor systems of \cas{} through a lack of other massive stars at their locations \citep[see methodology in][]{maund05, perets11, lyman14d}. The non-detection of other massive stars at the locations, alongside their diverse host galaxy type and galactocentric offset distributions, cannot be explained by massive progenitors. Indeed, even for the two examples investigated here that appear coincident with the hosts' disks, we find no evidence for underlying star formation at their location, ruling out very young progenitors. Although more moderately young progenitors cannot be ruled out for these examples when taken in isolation, since there is significant star formation in the vicinity, this interpretation does not fit with the class as a whole.

\begin{figure}
 \includegraphics[width=\columnwidth]{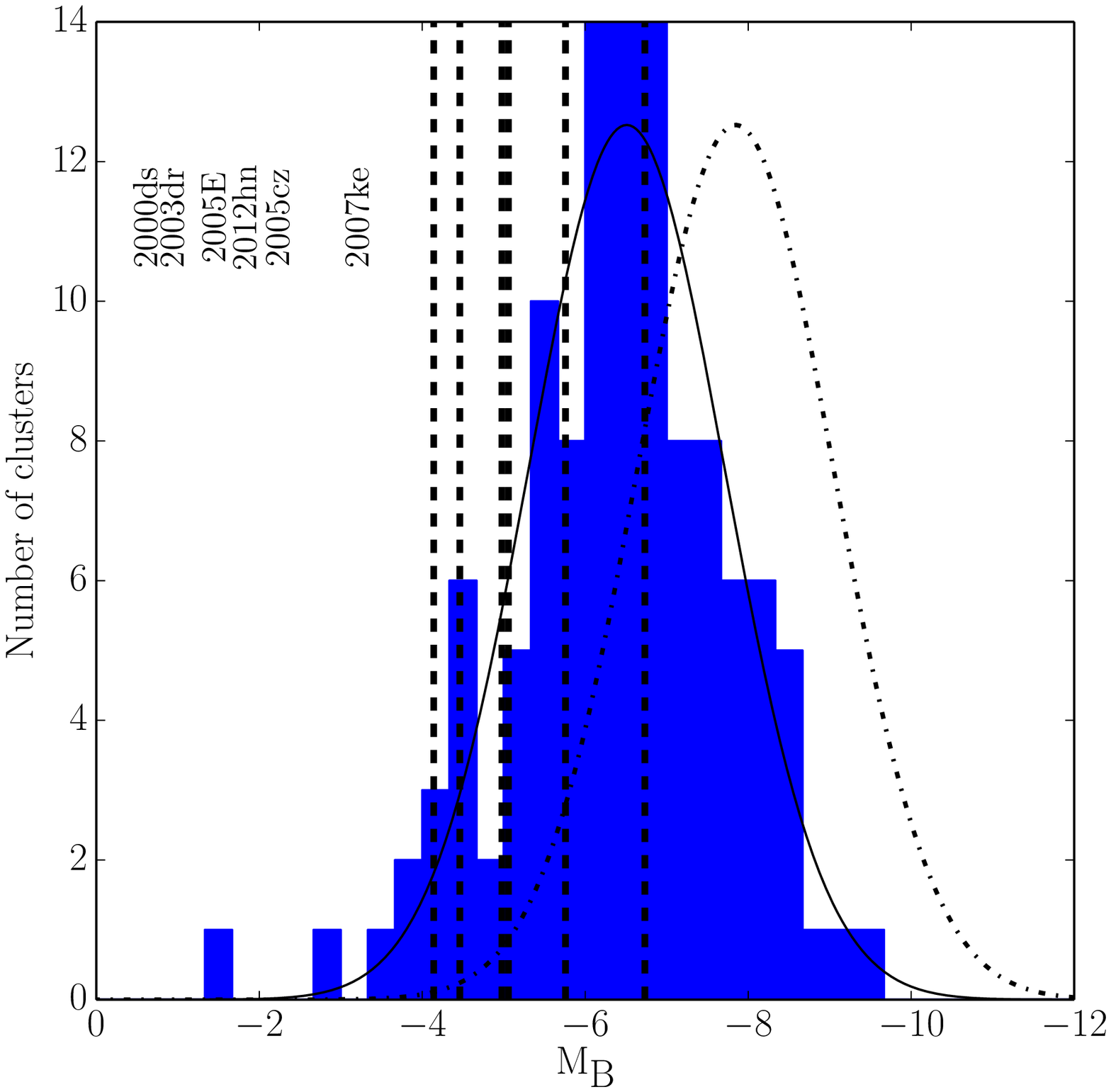}\\
 \includegraphics[width=\columnwidth]{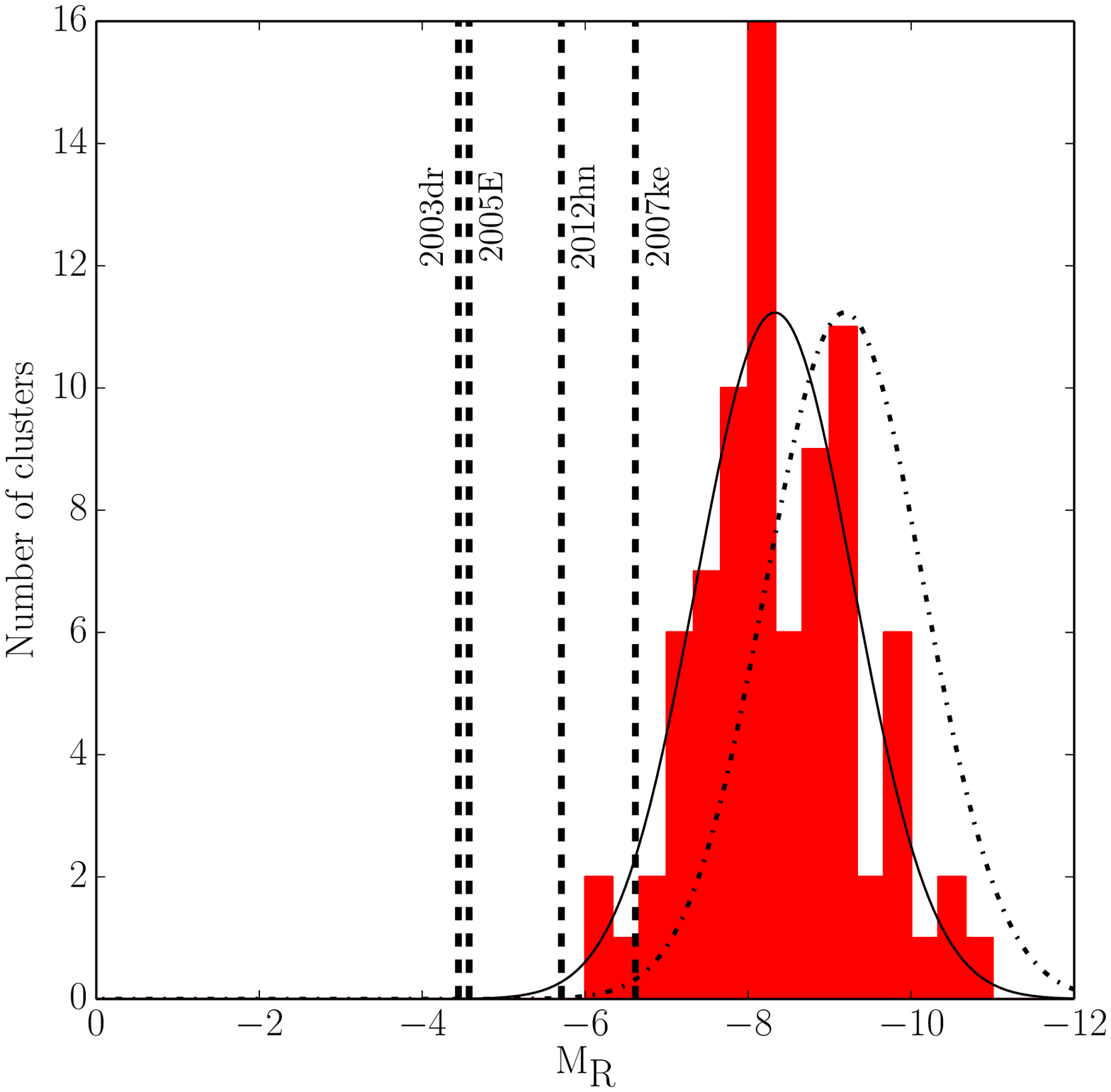}
 \caption{Deep limits at the locations of \cas{} (marked by vertical dashed lines) are inconsistent with the luminosity function of globular clusters (taken from \citealt{harris96}, 2010 edition). The labels for limits are offset in the {\em B}-band plot for clarity. A fitted normal distribution and the luminosity-weighted equivalent are shown by the black solid and dot-dashed lines, respectively. Distance moduli for the hosts of literature results \citep[see][and references therein]{lyman14d} were recalculated using the cosmology of this paper, based on recessional velocities from NED.} 
 \label{fig:gclf}
\end{figure}

Globular clusters as the host systems would present a solution to the extreme offset distribution of \cas{}, alongside their high stellar densities being conducive to the production of exotic close binary systems via dynamical interactions, the currently leading progenitor models. However the limits presented here rule out their presence at high significance, despite their luminosity distribution extending to quite faint absolute magnitudes. We compare our limits and those presented in \citet{lyman14d} to the Milky Way GC luminosity function \citep[][2010 edition]{harris96} in \cref{fig:gclf}. We assume for our \hst{} limits that F435W~$\simeq$~{\em B} and F606W~$\simeq$~{\em R}. After fitting the luminosity of GCs with a normal distribution, the {\em B} and {\em R}-band limits are inconsistent with the observed raw GC luminosity distribution at the 4.7 and 7.3 $\sigma$ level, respectively. When weighting by the clusters' luminosities, since one would expect the rate of events to increase with the luminosity (mass) of a cluster, these values become $\sim$7 and $> 8\sigma$. We note that statistically identical GC luminosity functions to that of the Milky Way were found for M31 and the nearby bright elliptical NGC~5128 \citep{rejkuba01}. Indeed, proposed variations in the GC luminosity functions between host galaxies appear to differ only at the $\sim 0.1-0.3$~mag level \citep{rejkuba12}, thus the specific choice of the GCLF is not affecting our conclusions.

The magnitude limits we present also rule out ultra-compact dwarf galaxies at the explosion sites of \cas{}. These have radii comparable to GCs but are somewhat brighter ($M_V \sim -11$ to $-14$, \citealt{evstigneeva08}). However, our point source limits may not be appropriate for other faint dwarf satellites populations, which can have radii of $\sim 200$~pc \citep[e.g.][]{belokurov07} and would thus be resolved in our imaging. We thus repeat our magnitude limit analysis for the remote sample but instead use a 200~pc aperture in order to estimate the limits on such an underlying dwarf galaxy population. Our limits in F435W are $-6.8$, $-6.8$ and $-7.1$~mag and in F606W are $-6.9$, $-6.5$ and $-7.3$~mag (for SNe 2003dr, 2005E and 2007ke, respectively).
Any dwarf galaxy population that may be underlying the explosion sites would thus have to be limited to a population that is $M \simgt{} -7$~mag. Such satellites are in the regime of ultra-faint dwarfs \citep{tolstoy09}. These systems are difficult to detect even around the Milky Way, requiring deep, wide-field and multi-colour surveys. The first two years of the Dark Energy Survey have found 16 new candidates and speculate $\sim$100 such systems associated with the Milky Way may be present over the whole sky \citep{bechtol15,drlica15}. However, the stellar masses of these systems are extremely low -- around 10$^3$~\msun{} (cf. Small Magellanic Cloud: $\sim 5\times10^8$~\msun{}).
Since these ultra-faint galaxies contain very little of the total stellar budget of dwarf galaxies in the local Universe, this would require very stringent progenitor formation constraints (such as an extreme metallicity cut-off for progenitor formation), which would have to act to restrict the formation of \ca{} progenitors in more luminous and massive satellite systems. Furthermore, the stellar densities in these ultra-faint galaxies are very low with consequently low rates of dynamical interactions to produce close binary system progenitors.

With these restrictions on the presence of faint underlying host systems, one may consider that these remote \cas{} are formed in the diffuse stellar halo surrounding the proximate brighter galaxies. As noted in \citet{lyman14d}, the haloes of galaxies are still too centrally concentrated to explain the offset distribution of the entire \ca{} sample, even when taking the extreme assumption of a progenitor population that comes {\em entirely} from the halo component of the galaxies.
To further compound this inconsistency, \citet{foley15} introduced a new member of the class, PTF~11kmb. This latest member is $\sim$~150~kpc from its putative host, meaning 5/13 members of the class are located greater than 20~kpc from their hosts in plane-of-sky alone.

In order to explain the peculiarities of their explosion sites, \citet{lyman14d} made the suggested mechanism of high-velocity binary systems as the progenitors of \cas{} based on a lack of underlying host systems. These new results expand this conclusion more broadly to the class as a whole by probing to similar limits for more distant examples through the use of \hst{}. The introduction of progenitors that travel large distances during their delay time to explosion can alleviate the problems of a lack of underlying host systems, and the consequent very low stellar densities at their explosion sites. Merging WD-NS binaries that have undergone kicks were suggested as a potential progenitor system \citep{lyman14d}. An initial study into the mergers of WD-NS as progenitors for the intermediate luminosity transients was made by \citet{metzger12}. Further investigations of these potential progenitors has been lacking, however a study into the mass transfer and merging of WD-NS in-spirals will be presented in Bobrick et al. (in preparation). 

The suggestion of high-velocity progenitor systems was backed up by a study of the kinematics of the transients relative to their host galaxies \citep{foley15}. It was found that \cas{} appear to be `expelled' from their hosts rather than on bound orbits (this is alternative evidence to that given above against \cas{} being hosted in ultra-faint dwarf galaxies, which our limits cannot probe).
In that study, interactions of binary or multiple star systems with the central super massive black hole (SMBH) of their hosts was suggested as the mechanism to create high-velocity progenitor systems. In this case the progenitor would be a high-velocity (and probably low-mass) WD-WD binary that would eventually merge. The offset angle distribution of \cas{} compared to the major axes of the putative hosts precludes very short-lived (post-kick) disk systems \citep{foley15}. However, a disk-borne population with a longer, or large spread in, merger time post-kick, as well as a nuclear-borne (SMBH interaction) progenitors, remain viable. 

The depth and resolution of these \hst{} data have also revealed clear signs of interaction and merging amongst the hosts. These data show NGC~1129 is the product of a recent major merger, as has been speculated previously, and show a conspicuous tidal feature around NGC~5714, which is likely to be the result of an ongoing minor merger. The high occurrence of disturbed and/or merging systems was noted by \citet{foley15}, with the results of this study adding further support.  Unfortunately, membership of the class remains very low (13). Should the observed trends in \ca{} hosts continue as new members are found, the biased production of \cas{} in such systems must be explained by any proposed progenitor system.

\section{Summary}
\label{sect:summary}

We have presented new \hst{} observations of the hosts and locations of five nearby \cas{}. Two of these appear in line-of-sight with the disks of late type galaxies, and three are `remote' examples, lying well outside the bulk of their putative hosts' stellar light distribution. 

For the three remote events, SNe 2003dr, 2005E and 2007ke, we obtained deep imaging with ACS in F435W and F606W. We found no detected sources underlying the locations of these transients, ruling out the presence of massive stars, dwarf galaxies and globular clusters at these locations. The lack of faint host systems and the large galactocentric offsets indicate the stellar density is extremely low at these locations. This makes their locations difficult to explain with traditional progenitor suggestions, whose distributions would be expected to trace the stellar light of the hosts.

The two examples coincident with a galactic disk, SNe 2001co and 2003dg, were observed with WFC3 in F438W and F606W, with a narrowband image in in F665N, centered on \Ha{} at the redshift of their hosts. In each case the transient's location appears close to regions of star-formation. Given the morphology of the hosts ($\sim$Sbc), the presence of some degree of nearby star formation is unsurprising. However, it is telling that there locations are not directly overlaid on regions of \Ha{} emission, as may be expected for very young progenitor systems. Furthermore, the lack of distinguishable underlying sources at their locations (e.g. as would be expected for a GC origin) is in agreement with the findings for the remote sample, although in these cases there is clearly an underlying stellar population from the galactic disk. 

With these new data, deep observations have now been taken at the location of nine of the ten known \cas{} within 100~Mpc. A common theme for the remote examples is the lack of any underlying compact stellar systems at the locations of the explosions. Those within the disk of their hosts also lack signatures of an obvious underlying birth place beyond the diffuse stellar population (e.g. SNe 2000ds: \citealt{maund05}, 2005cz: \citealt{perets11}, 2003H: \citealt{lyman14d}, 2001co and 2003dg: this study). High-velocity systems, which travel significant distances from their birth sites appear to alleviate many of the problems (i.e. remote locations and kinematics of the transients around their hosts) of other progenitor suggestions, and could provide an explanation for these peculiar explosions \citep{lyman14d,foley15}.

An interesting feature amongst \ca{} hosts is the apparent prevalence of disturbed or merging systems. This is particularly relevant for progenitor models including SMBH-interactions of the progenitor systems -- mergers are likely to produce binary SMBH systems, increasing the rate of SMBH-interactions \citep{foley15}. Our imaging has shown firm evidence of relatively recent merging activity in two hosts (NGC~5714 and NGC~1129). As new members of the class emerge, it will be prudent to further test this apparent bias of \ca{} production in disturbed and merging systems.

\section{Acknowledgements}

We thank Joe Anderson and the Carnegie Supernova Project for providing imaging of SN~2005E. Llu\'{i}s Galbany and the CALIFA survey are thanked for allowing us to inspect the CALIFA data cube of NGC~5714 prior to its public release in Data Release 3. JDL and AJL acknowledge support from the UK Science and Technology Facilities Council (grant ID ST/I001719/1). RPC was supported by the Swedish Research Council (grant 2012-2254).

Based on observations made with the NASA/ESA Hubble Space Telescope, obtained at the Space Telescope Science Institute, which is operated by the Association of Universities for Research in Astronomy, Inc., under NASA contract NAS 5-26555. These observations are associated with program GO13698.

\newcommand{\araa}{ARA\&A}   \newcommand{\aap}{A\&A}
\newcommand{\aj}{AJ}         \newcommand{\apj}{ApJ}
\newcommand{\apjl}{ApJ}      \newcommand{\apjs}{ApJS}
\newcommand{\mnras}{MNRAS}   \newcommand{\nat}{Nature}
\newcommand{\pasj}{PASJ}     \newcommand{\pasp}{PASP}
\newcommand{\procspie}{Proc.\ SPIE} \newcommand{\physrep}{Phys. Rep.}
\newcommand{\apss}{APSS}     \newcommand{\pasa}{PASA}
\newcommand{\solphys}{Sol. Phys.}
\newcommand{\actaa}{Acta Astronom}
\newcommand{\aaps}{A\&A Supp}
\newcommand{\iaucirc}{IAU Circular}
\bibliographystyle{mn2e}
\bibliography{/home/jdl/references}

\label{lastpage}
\end{document}